\documentclass[review]{elsarticle}

\usepackage{lineno,hyperref}
\usepackage{multirow}
\usepackage{caption}
\usepackage{subcaption}
\usepackage{mathtools}
\usepackage{algorithm}
\usepackage{algorithmic}
\usepackage{cleveref}

\newtheorem{definition}{Definition}

\usepackage{lipsum}
\makeatletter
\def\ps@pprintTitle{%
 \let\@oddhead\@empty
 \let\@evenhead\@empty
 \def\@oddfoot{}%
 \let\@evenfoot\@oddfoot}
\makeatother

\begin{document}

\begin{frontmatter}

\title{Bowlership: Examining the Existence of Bowler Synergies in Cricket}

\author{Praharsh Nanavati}
\address{Department of Data Science and Engineering}
\address{IISER, Bhopal}

\author{Amit A. Nanavati}
\address{School of Engineering and Applied Science}
\address{Ahmedabad University, Ahmedabad}




\begin{abstract}
Player synergies are a salient feature of team sports. In the team game of
cricket, player synergies may be reflected in batting partnerships. 
Batting partnerships have been analysed extensively. In this paper, we 
introduce and precisely define bowling partnerships. 
We explain their importance, and analyse ball-by-ball data from three formats
of the game: 2,034 one-day international matches, 634 Test matches and 1,432
Twenty-20 international matches, in order to find such bowling partnerships
(``bowlerships"). We find that bowlerships exist. Further, we construct
bowlership networks based on these pairwise synergies. We assert that these
bowlership networks can be analysed for team selection before a match, and
making bowling changes during the match. We present Algorithm 
{\textbf{\slshape bowler-select}} that selects a team based on \textit {bowlerships}.

\end{abstract}


\end{frontmatter}


\section{Introduction}

Team sports are all about player synergies; deciding when to let your partner take the shot in Tennis doubles or anticipating the next pass in Football. This is why often the best single's champions are not necessarily the best doubles' champions in Tennis, and the team with more star players does not necessarily win the Football match. 

Cricket is a team sport played in most Commonwealth countries. There are various formats of the game varying from five-day long matches to 3-hour matches. Batsmen and Bowlers in Cricket are traditionally ranked according to their batting and bowling averages respectively~\cite{mukherjee2014quantifying}. 
Batsmen bat in pairs, and the pair is known as a partnership. A partnership continues batting until one of the batsmen is dismissed. It has long been believed that synergies exist between effective batting partners. Some batting pairs are part of legend. For example, Matthew Hayden and Justin Langer of Australia and Desmond Haynes and Gordon Greenidge of the West Indies are mentioned by Wisden~\cite{wisdenbat}. 
In their paper~\cite{valero2012investigation}, the authors investigate the importance of batting partnerships in Test and one-day cricket with respect to improved performance.  However, based on their statistical analyses, the authors conclude that synergies in opening partnerships may be considered a sporting myth.

In this paper, we investigate bowler pairs. We introduce and precisely define a bowling partnership. Bowlers bowl in pairs from opposite ends of the field alternately. It is often the case that a pair of bowlers bowl several consecutive overs alternately. We define such pairs of bowlers a \textit{bowling pair}. Are some bowling pairs more effective than others? Are they effective in terms of saving runs or taking wickets?

Wisden also lists famous bowling partnerships~\cite{wisdenbowl}. However, the bowling 
partnership list contains pairs of bowlers who bowled in the same match (but not necessarily together). 

Bowlers have two goals: saving runs and taking wickets. Commentators sometimes anecdotally observe that when one of the bowlers is economical (saving runs), the other bowler is targeted by the batsmen leading them to make errors and give away wickets. This raises the question whether two bowlers are more effective as a pair with each other -- whether in saving runs or taking wickets or both.

\begin{definition}
A bowler's economy rate is the average number of runs he/she has conceded per over bowled. The lower the economy rate is, the better the bowler is performing.
\end{definition}

\begin{definition}
A bowler's hitrate is defined as the average number of wickets he/she has taken per over bowled. The higher the hitrate, the better the bowler is performing.
\end{definition}

Deciding which bowler should bowl from which end and when is a crucial decision to be made by the captain. As with other things, bowler (team) selection (before the match) and deciding bowling changes (during the match) takes skill and experience. Such decisions depend upon a number of complex factors including pitch conditions and assessing the competitor team's strengths and weaknesses. If effective bowling partnerships do in fact exist, then this could inform the
strategies for team selection as well as bowling changes.

Bowlers are typically assessed on two aspects -- the wickets they take (the more the better) and the runs they concede (the less the better). While the former matters the most in Test cricket (5-day matches), in the limited overs versions (one-day internationals and T20 matches) the latter matters too. The Bowling strike rate is defined for a bowler as the average number of balls bowled per wicket taken~\cite{strikeratebowl}, and  the economy rate is the average number of runs they have conceded per over bowled~\cite{economyratebowl}.
Therefore a bowling pair can be deemed effective (synergistic) if the pair together saves
more runs and/or takes more wickets. This could happen in three ways: (i) both the bowlers have 
improved economies, (ii) both the bowlers take more wickets, (iii) one of them has an improved economy while the other gets more wickets. 

Our analysis involves a comparison of the performance of a bowler together with his partners to
investigate if the performance of the bowler with a given partner is statistically superior to his performance with the rest. The findings of such analysis can inform bowler selection and bowling changes. 



We analysed data across all formats of the game --- Test Cricket, ODIs and T20Is. 
In Section 2, we define and explore these bowling partnerships, and investigate if there are any statistically significant partnerships. In section 3, we present results from the 3 formats of the game and compare them. Some discussion and concluding remarks are then provided in Section 4.

\section{Methodology}

We acquired ball-by-ball data of Men’s Test Matches, One-day internationals and T20 Internationals in separate files in the YAML format from cricsheet.org~\cite{Cricshee33:online}. These include metadata like venue, date, participating teams, toss details including striker, non-striker, bowler, runs scored on each ball, wickets taken (if any), type of dismissal, extras information and outcome of the game.

 \begin{figure}
     \centering
          \begin{subfigure}[b]{0.47\textwidth}
          \centering
          \includegraphics[width=\textwidth]{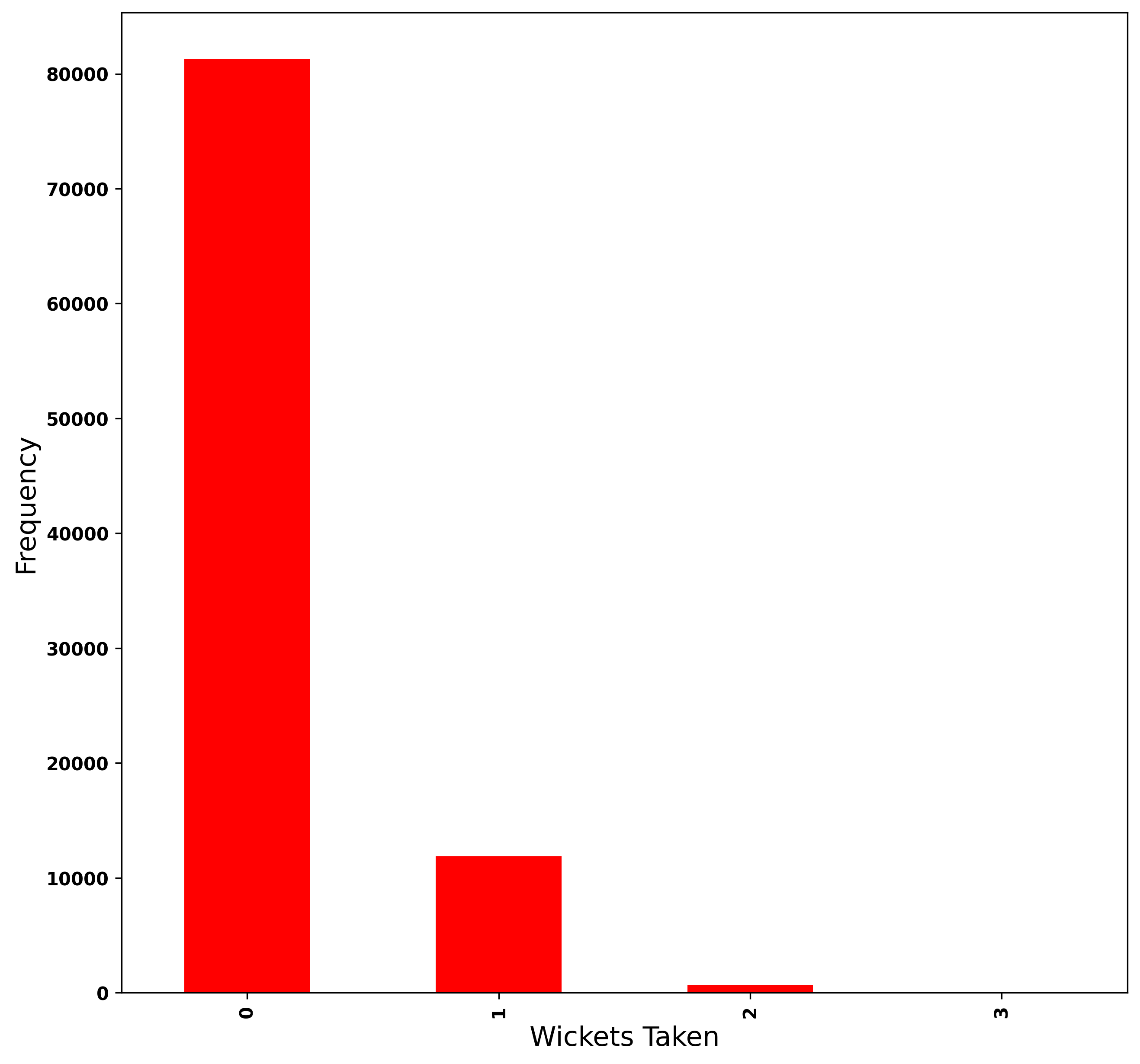}
          \caption{ODI wickets}
          \label{sfig:ODIwickets}
          \end{subfigure}
          \hfill
          \begin{subfigure}[b]{0.47\textwidth}
          \centering
          \includegraphics[width=\textwidth]{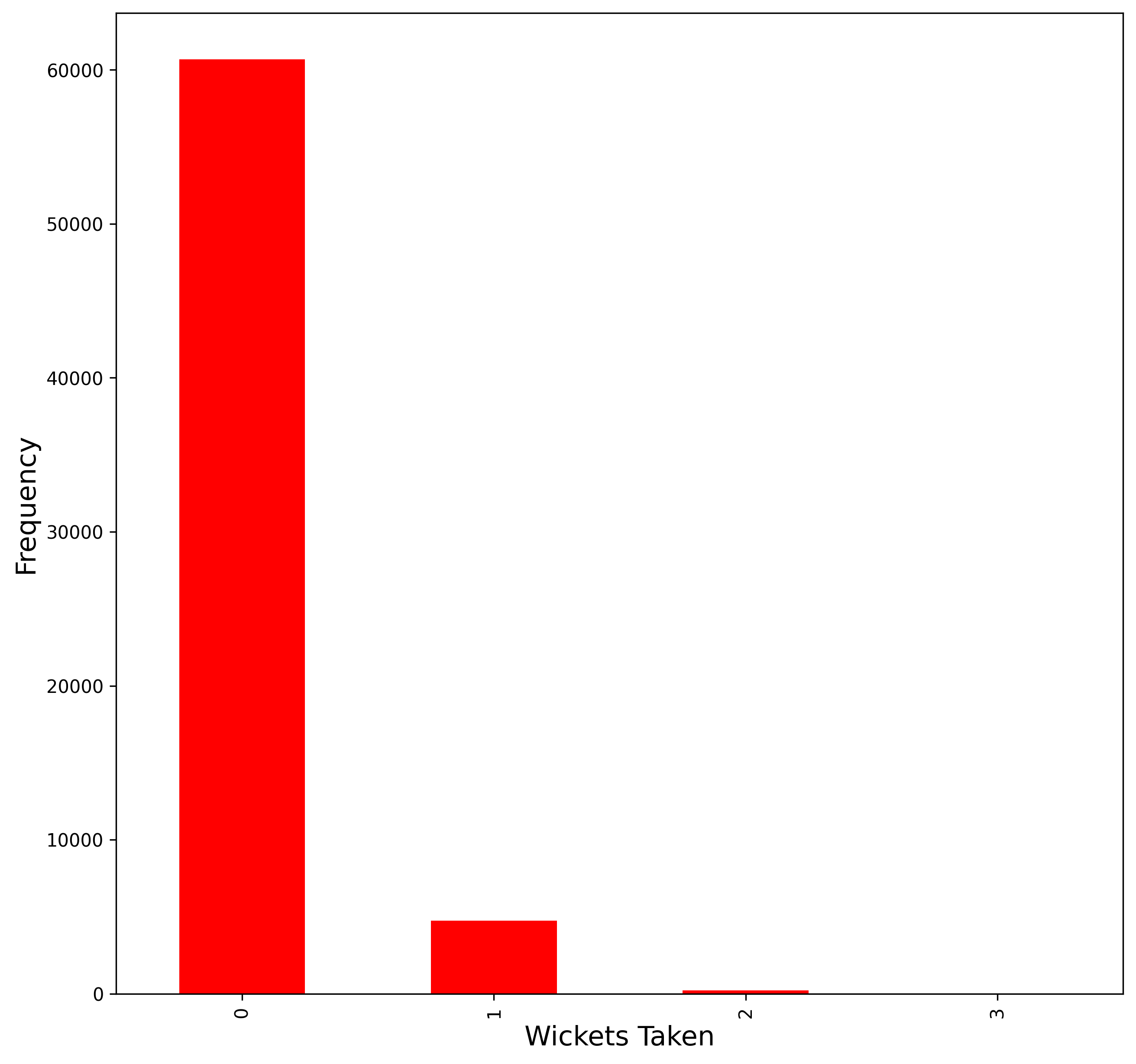}
          \caption{Test wickets}
          \label{sfig:Testwickets}
          \end{subfigure}
          \hfill \\
         \begin{subfigure}[b]{0.47\textwidth}
          \centering
          \includegraphics[width=\textwidth]{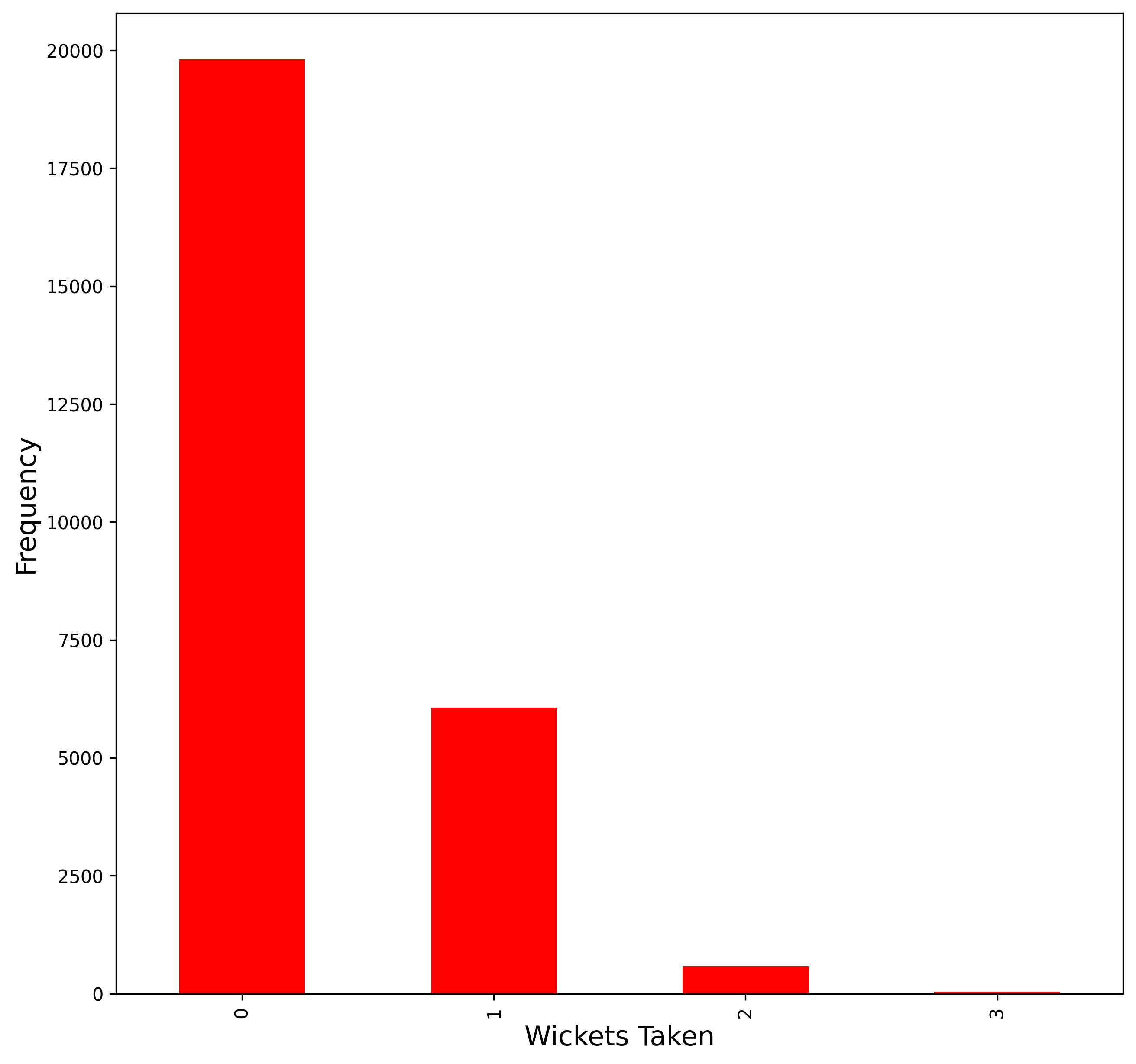}
          \caption{T20I wickets}
          \label{sfig:T20wickets}
          \end{subfigure}
          \hfill
     \caption{This figure shows that no wicket falls in most overs across all formats. The falling of a wicket is a rare event. Taking more than one wicket in an over is extremely uncommon.}
    \label{fig:wkts}
 \end{figure}


We analysed 2,034 ODIs, 634 Test matches and 1,432 T20Is. 
There were a total of 1148 ODI bowlers, 495 test bowlers and 1518 T20I bowlers. 
Figure~\ref{fig:wkts} shows the number of overs with the number of wickets taken across all formats. Note that most overs in all formats did not see any player getting out. In general, the falling of a wicket is a rare event. This is quite unlike scoring runs.

 \begin{figure}
     \centering
          \begin{subfigure}[b]{0.45\textwidth}
          \centering
          \includegraphics[width=\textwidth]{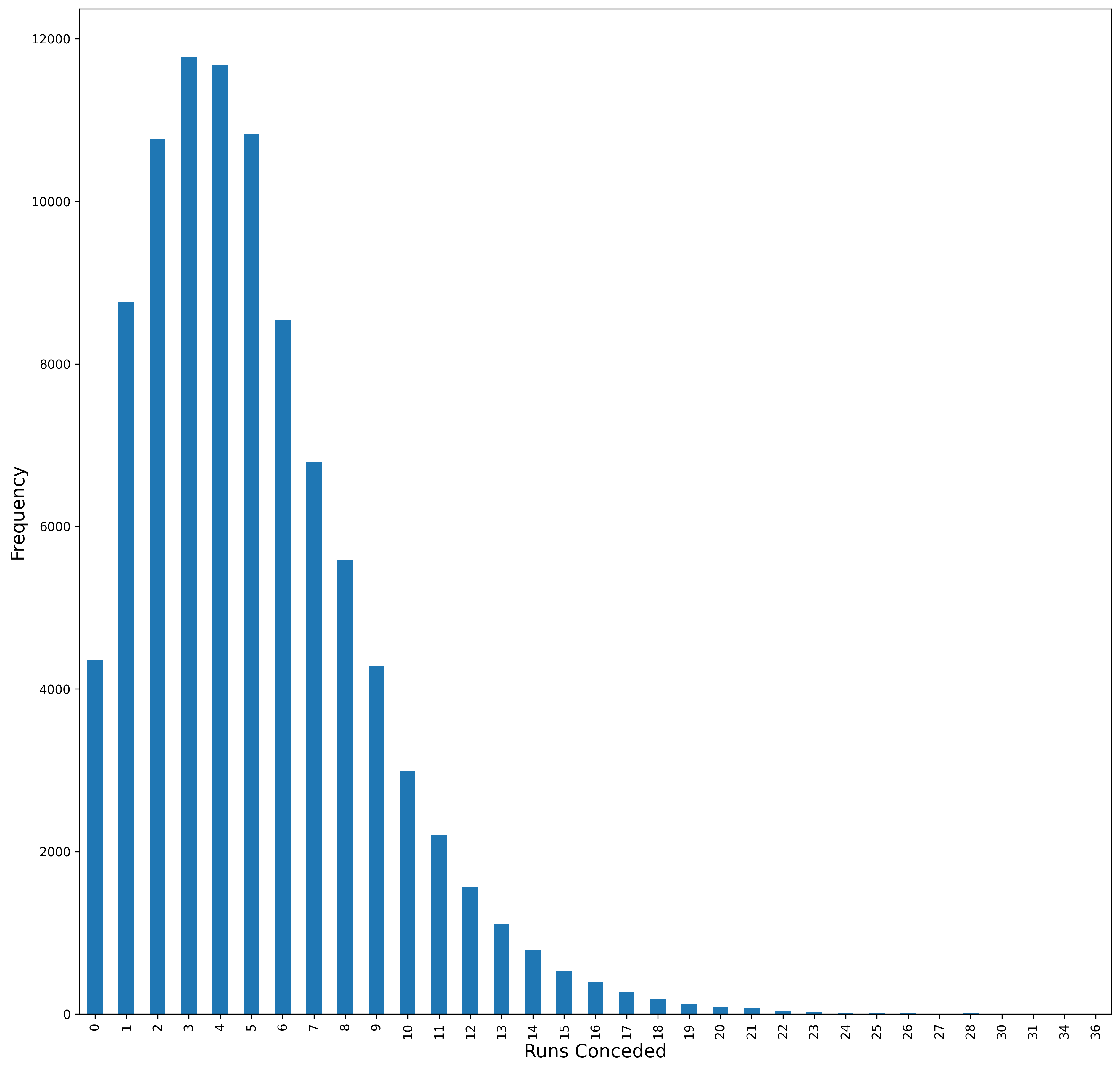}
          \caption{ODI Economy}
          \label{sfig:ODIeconomy}
          \end{subfigure}
          \hfill
          \begin{subfigure}[b]{0.45\textwidth}
          \centering
          \includegraphics[width=\textwidth]{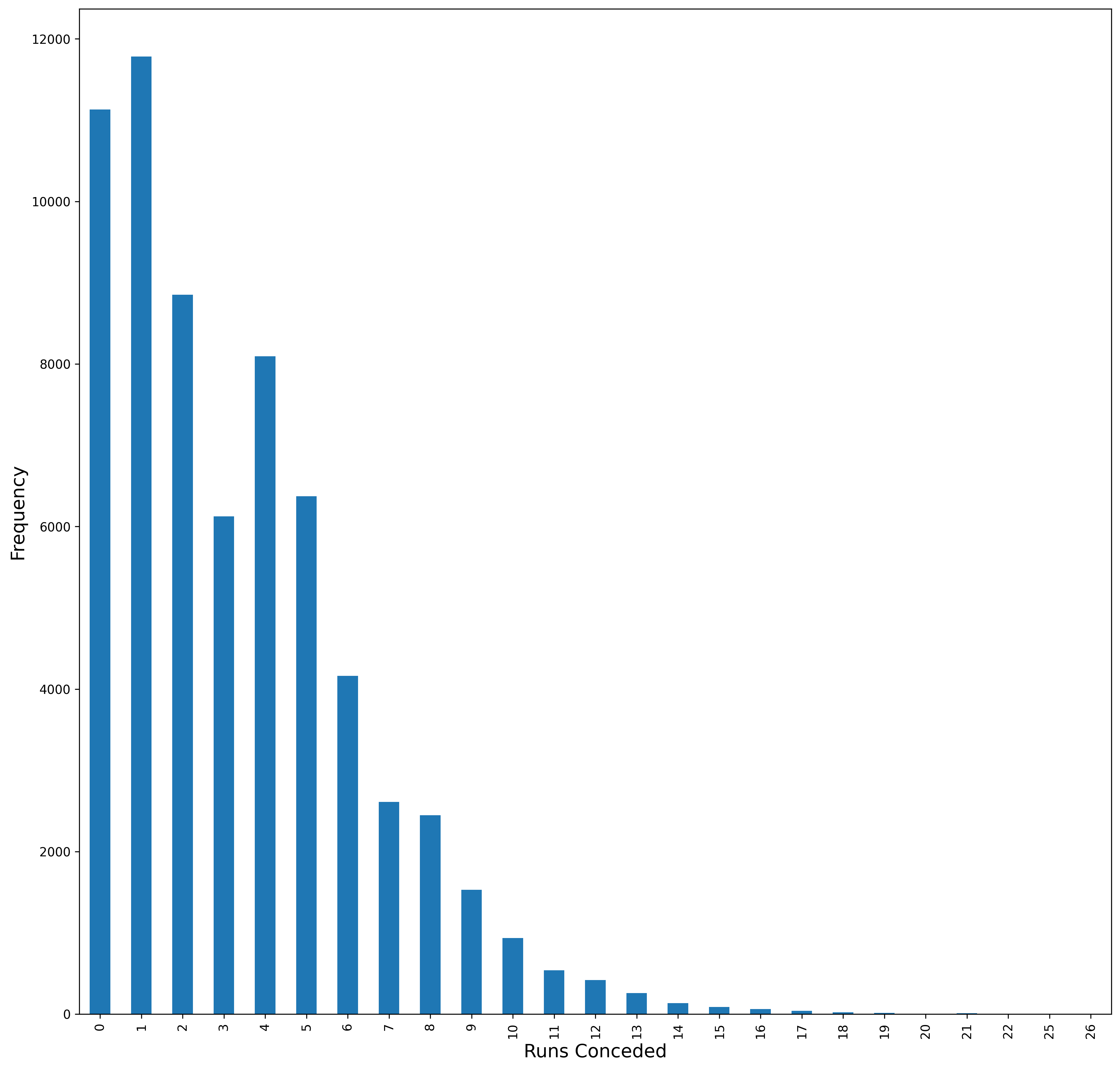}
          \caption{Test Economy}
          \label{sfig:Testeconomy}
          \end{subfigure}
          \hfill \\
         \begin{subfigure}[b]{0.45\textwidth}
          \centering
          \includegraphics[width=\textwidth]{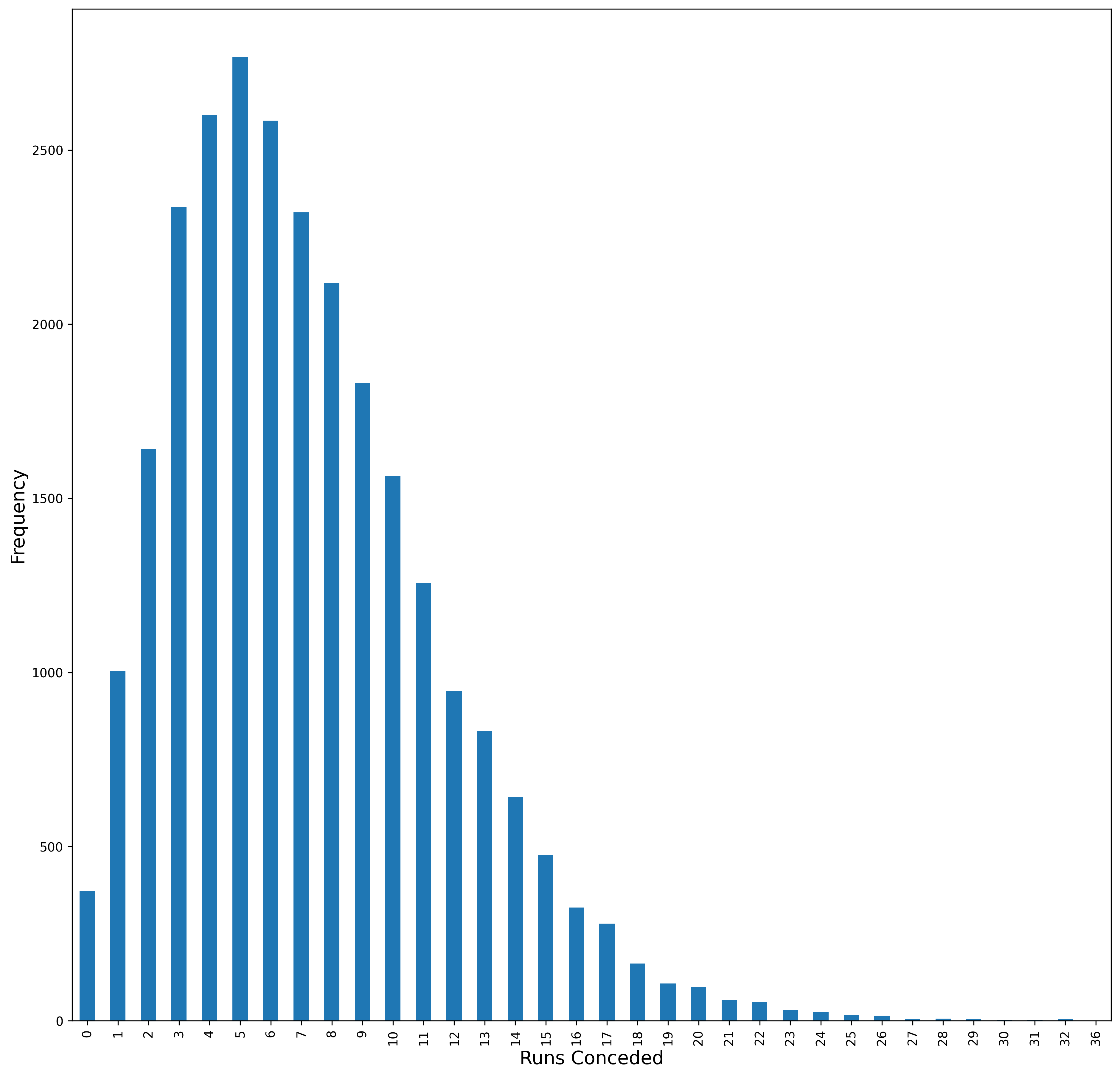}
          \caption{T20I Economy}
          \label{sfig:T20economy}
          \end{subfigure}
          \hfill
     \caption{This figure shows the frequency distribution of runs conceded per over for all matches for ODIs(a), Test Matches(b) and T20Is(c). In ODIs and T20Is, It is uncommon to concede too few or too many runs in an over. In test matches, the number of low scoring overs is much higher. The shorter the format, the higher the likelihood of conceding more runs in an over.}
    \label{fig:economy}
 \end{figure}


\begin{figure}
    \centering
    \includegraphics[width=\textwidth]{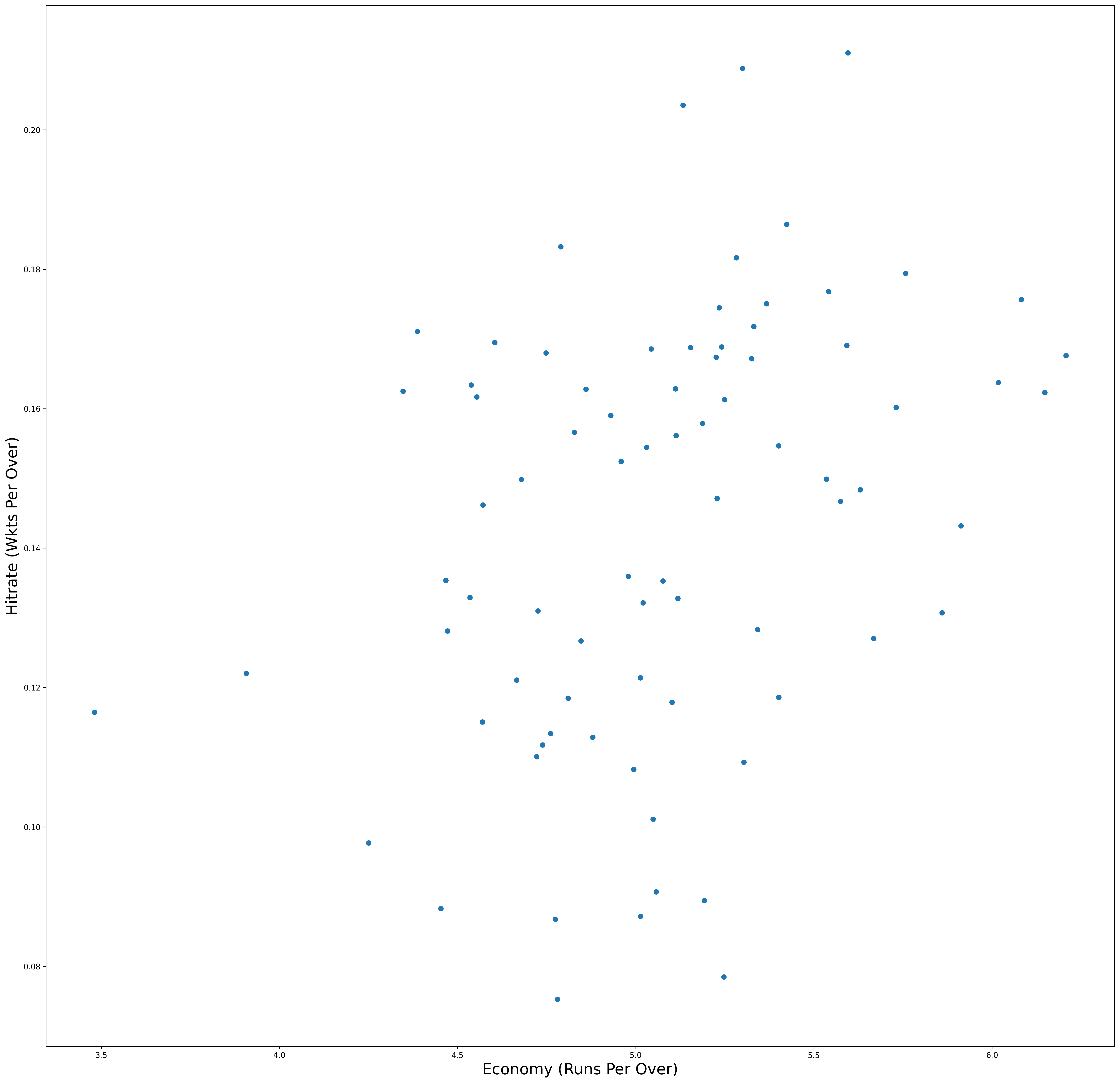}
    \caption{ODI Hitrate vs. Economy: This scatter plot show the economy on the X-axis along with their respective hitrates on the Y-axis for all the bowlers. 
    Bowlers to the top left are good at both traits -- conceding fewer runs and taking more wickets whereas bowlers to the bottom right are conceding the most runs and taking the fewest wickets per over.}
    \label{fig:ODIscatter}
\end{figure}
 
\begin{figure}
    \centering
    \includegraphics[width=\textwidth]{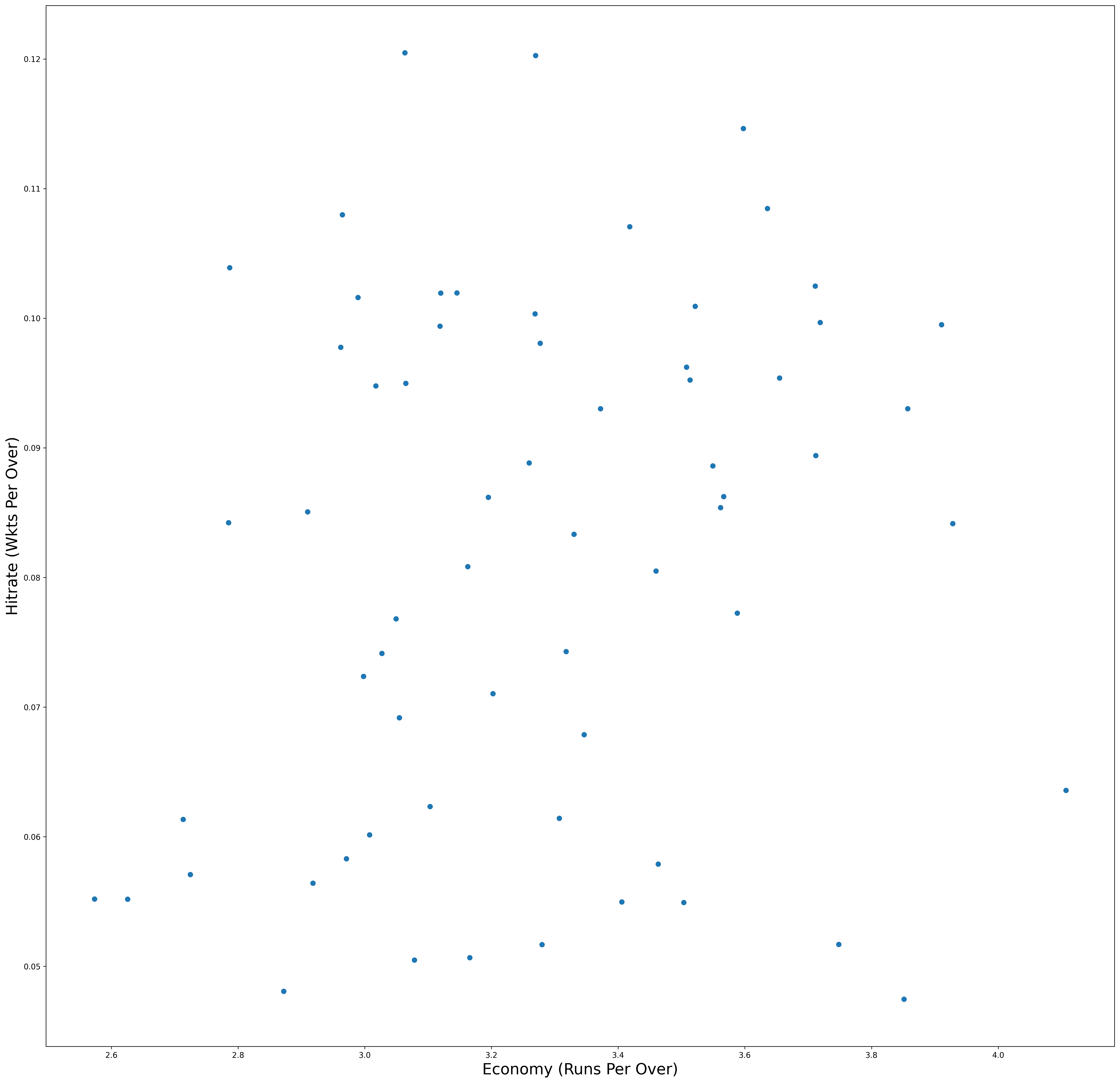}
    \caption{Test Hitrate vs. Economy: This scatter plot show the economy on the X-axis along with their respective hitrates on the Y-axis for all the bowlers. 
    Bowlers to the top left are good at both traits -- conceding fewer runs and taking more wickets whereas bowlers to the bottom right are conceding the most runs and taking the fewest wickets per over.}
    \label{fig:Testscatter}
\end{figure}

\begin{figure}
    \centering
    \includegraphics[width=\textwidth]{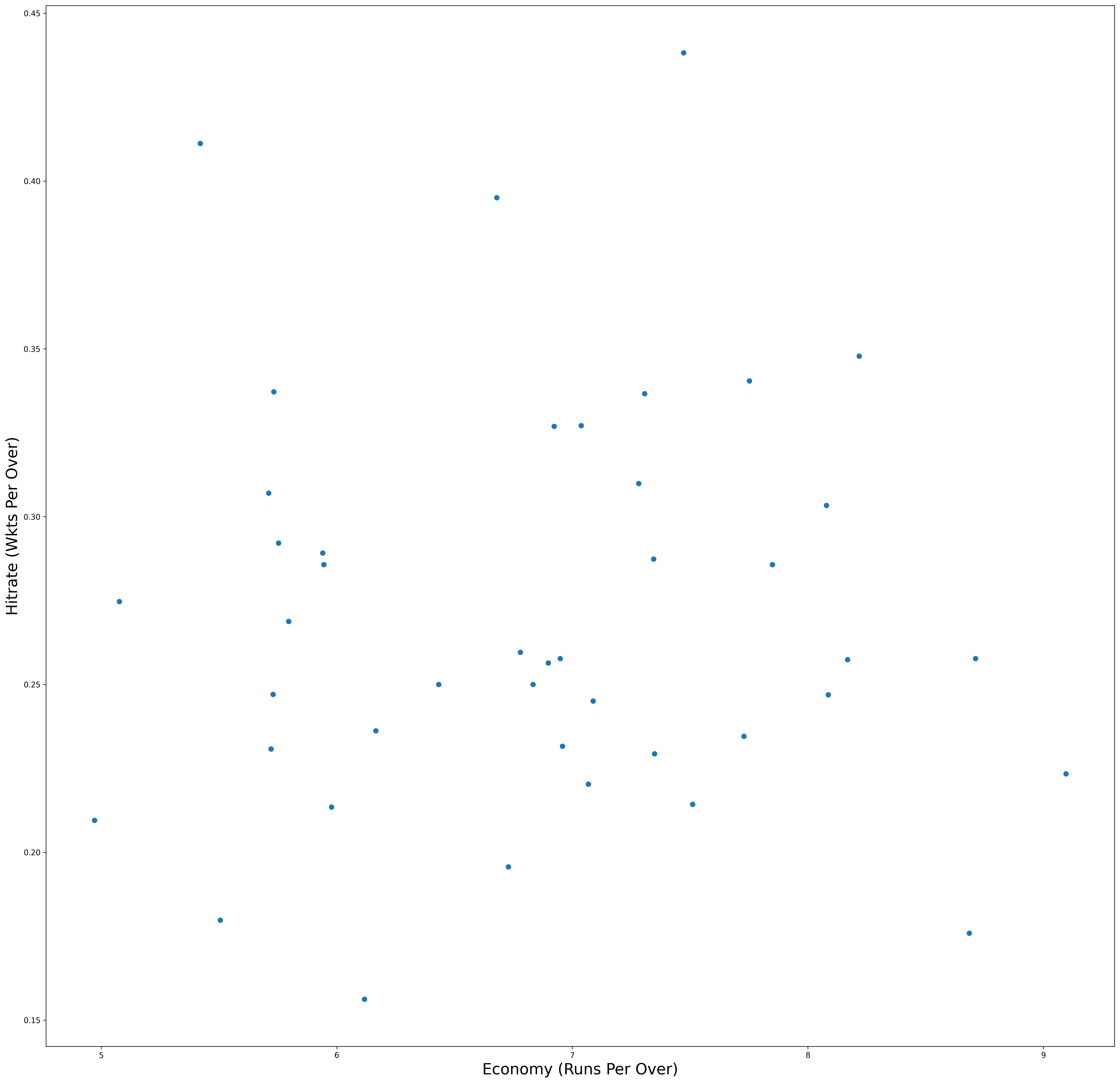}
    \caption{T20I Hitrate vs. Economy:This scatter plot show the economy on the X-axis along with their respective hitrates on the Y-axis for all the bowlers. 
    Bowlers to the top left are good at both traits -- conceding fewer runs and taking more wickets whereas bowlers to the bottom right are conceding the most runs and taking the fewest wickets per over.}
    \label{fig:T20scatter}
\end{figure}

Figure~\ref{fig:economy} shows the distribution of runs conceded by all bowlers across all overs. Notice that conceding runs appears to be a normal curve. It is uncommon to concede too few or too many runs in an over.
~\Cref{fig:ODIscatter,fig:Testscatter,fig:T20scatter} shows the Economy v Hitrate plots indicating the trends of all the bowlers in each format.

\begin{table}[h]
    \centering
    \begin{tabular}{|c||c|c||c|c||c|c|} \hline
    \multirow{2}{*}{Statistical Test }           & \multicolumn{2}{|c||}{ODIs} & \multicolumn{2}{|c||}{Tests} & \multicolumn{2}{|c|}{T20Is} \\ \cline{2-7}
        (Normality) & Fail & Pass &  Fail & Pass & Fail & Pass \\ \hline
    \multirow{2}{*}{Chi-square} & 578  & 393  & 295  & 129  & 232  & 609  \\
                                &(60\%)&(40\%)&(70\%)&(30\%)&(28\%)&  (72\%) \\ \hline
    \multirow{2}{*}{Shapiro-Wilk}    & 674 & 297 & 360 &  64 & 316 & 525 \\ 
       & (71\%)& (29\%)& (85\%)&  (15\%)& (38\%)& (62\%)\\ \hline
    \multirow{2}{*}{Anderson-Darling}& 785 & 186 & 390 &  34 & 463 & 378 \\ 
     & (81\%)& (19\%)& (92\%)&  (8\%) & (55\%)& (45\%)\\ \hline
    \end{tabular}
    \caption{The results of 3 statistical tests of Normality. Note that the test of normality failed several data points.}
    \label{tab:norm}
\end{table}

To measure a bowler's performance, Croucher~\cite{croucher2000player} defines the bowling index as:
$$ \textit{Bowling Index} = \textit{Bowling average} \times \textit{Bowling Strikerate} $$
where, bowling average is the the number of runs conceded by a bowler per wicket taken and bowling strikerate is the average number of balls bowled for every wicket taken. However, a bowler may be considered successful if he takes wickets and/or gives away few runs and hence we 
prefer to keep the Economy and the Hitrate separate.

We need to analyse an individual bowler's economy as well as hitrate. For each bowler,
we plotted the runs conceded per over and checked if these distributions were normal. Table~\ref{tab:norm} summarises the results. We conducted three normality tests: Chi-square test~\cite{d1971omnibus}, Shapiro-Wilk test~\cite{shapiro1965analysis} and Anderson-Darling test~\cite{stephens1974edf}. Except in the case of T20Is for the former two tests, more bowlers fail than pass the test. Therefore, for most bowlers, the distribution is not normal. Since the falling of a wicket is a rare event, the distribution of wickets per over for each bowler is not normal either.

\begin{definition}
A pair of bowlers constitute a {\em bowling pair} at individual threshold $t_i$ and pairing-threshold $t_p$ iff: 
(a) each bowler has bowled at least $t_i$ overs in his career, and 
(b) together they bowl at least $t_p$ consecutive overs alternately over all the matches.
\end{definition}



In order to exclude trivial bowling pairs, we set the following conditions:
\begin{description}
    \item [T1.] (for $t_i$) The individual bowlers in a bowling pair should have bowled at least 300 overs (in Tests), 300 overs (in ODIs) and 80 overs (in T20Is) throughout the span of their careers.
    \item [T2.] (for $t_p$) In order for a pair of bowlers to be considered a bowling pair, we set the pairing-threshold -- the number of consecutive overs that they should have bowled alternately -- to 60 (in Tests), 60 (in ODIs) and 16 (in T20Is). 
\end{description}

Based on condition T1, we found:
\begin{itemize}
\item 64 Test bowlers (out of 495) who have bowled at least 300 overs.
\item 80 ODI bowlers (out of 1148) who have bowled at least 300 overs.
\item 45 T20I bowlers (out of 1518) who have bowled at least 80 overs.
\end{itemize}

Among the bowlers who individually satisfied T1, based on the additional condition T2, we got:
\begin{itemize}
\item 81 Test bowler pairs who have together bowled at least 60 overs.
\item 41 ODI bowler pairs who have together bowled at least 60 overs.
\item 18 T20I bowler pairs who have together bowled at least 16 overs.
\end{itemize}



While these numbers appear to have been arbitrarily chosen, the basic rationale is to ensure that an individual bowler has bowled enough overs and the choice of the pairing-threshold is such that it allows a bowler to have a few potential bowling pairs (5 with the above values). The analysis detailed in this section could easily be conducted by choosing other values to obtain corresponding results. We discuss this further in Section~\ref{sec:concl}.

We need to compare the performance of an individual bowler with the bowling pairs that the bowler is a part of. Since the individual distributions are not normal, we will use a non-parametric test, the Mann-Whitney test to do the comparisons.

\subsection{Mann-Whitney U Test}

The Mann-Whitney U test is a nonparametric test of the null hypothesis that, for randomly selected values X and Y from two populations, the probability of X being greater than Y is equal to the probability of Y being greater than X~\cite{mwtest}. Since not all bowlers' economy rates or wickets per over follow a normal distribution,
we use this test to compare an individual bowler's performance with a paired performance of the same bowler with a partner. 

We consider the set of all the overs that bowler $A$ bowls with a particular partner $B$ (bowlership set) and the set of all the overs bowled by the bowler $A$ (individual set). The number of runs conceded in each over is considered for the Mann-Whitney U test. We conduct three tests: 
\begin{enumerate}
    \item ``greater" test: \\ H0: Individual Economy better than or same as the Bowlership Economy.
    \item ``two-sided" test:\\ H0: Individual Economy is same as the Bowlership Economy.
    \item ``less" test: \\ H0: Individual Economy worse or same as the Bowlership Economy.
\end{enumerate}

If the first two  tests fail, then the two null hypotheses can be rejected and we can conclude that the bowlership pair performs better than the individual. In this case, we say that a {\it positive bowlership} exists from $A$ to $B$. This relationship is not symmetric. Bowler $A$ may bowl better with a Bowler $B$, while the opposite need not be true. If the last two  tests fail, then the two null hypotheses can be rejected and we can conclude that the bowlership pair performs worse than the individual. In this case, we say that a {\it negative bowlership} exists from $A$ to $B$.
We conduct similar tests for Bowlership Hitrates.

\subsection{Bowlership Networks}

\begin{figure}
    \centering
          \begin{subfigure}[b]{\textwidth}
         \centering
          \includegraphics[width=\textwidth]{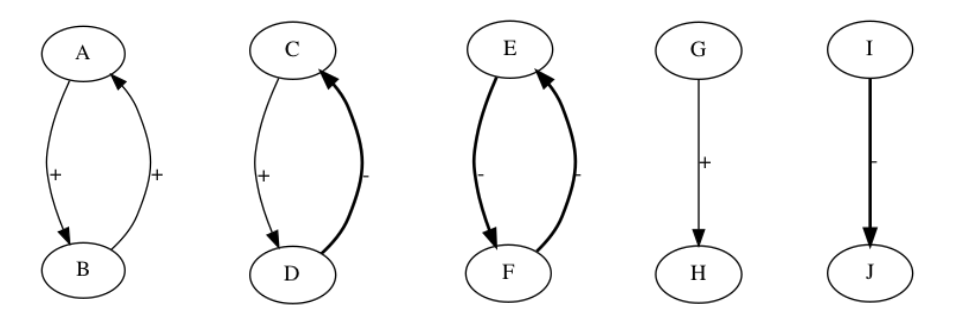}
          \caption{Examples of directed signed graphs on a pair of vertices.}
          \label{sfig:cases-a}
          \end{subfigure}
          \hfill \\
          \begin{subfigure}[b]{\textwidth}
         \centering
          \includegraphics[width=\textwidth]{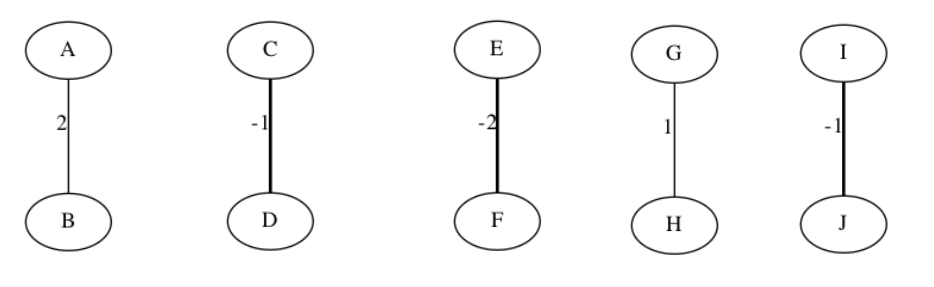}
          \caption{Corresponding weighted undirected versions.}
          \label{sfig:cases-b}
          \end{subfigure}
          \hfill 
    \caption{Conversion of directed signed graph into a weighted undirected graph.}
    \label{fig:motifs}
\end{figure}

We can construct a directed signed graph $G_d=(V,\overrightarrow{E})$  where $V$ is the set of bowlers, and we draw 
a {\it positive directed} edge from a bowler $A$ to $B$ if $A$ bowls better with $B$ and a
{\it negative directed} edge if $A$ bowls worse with $B$. Figure~\ref{sfig:cases-a} indicates the various cases between a pair of bowlers. 

We can analyse these graphs to suggest bowling changes during the match as well as team (bowler) selection before the match.
The basic idea is to select a set of bowlers with as many positive bowlership pairs as possible to give the captain maximum flexibility in making bowling changes during the match. 

Towards this, we first convert the directed signed graph $G_d$ into an undirected weighted graph $G_u$. 
Each signed directed edge is replaced by an undirected weighted edge as depicted in the exhaustive combinations shown in Figure~\ref{fig:motifs}.    
In particular, note that while the directed version for nodes $C,D$ as well as nodes $I,J$ are different, they result in the same weighted undirected
graph. This is because a negative edge from $A$ to $B$ nullifies the effect of a positive edge from $B$ to $A$, since $A$ and $B$ are essentially incompatible. After this transformation, the weight of an edge indicates the strength of bowlership between the endpoints. 

\begin{definition}
A subgraph $S$ of the undirected weighted graph $G_u$ (which may contain negative edges), has an associated {\em average weighted degree}
$$W(S) = {\sum w(S) \over |S|}$$
where $w(S)$ is the sum of weights of all edges induced by $S$ and $|S|$ is the number of vertices in $S$. 
\end{definition}

\begin{algorithm}
\caption{{\textbf {\slshape create-weighted-graph}}}
\label{alg:bsel}
\begin{algorithmic}[1]
\REQUIRE directed signed graph $G_d$ 
\ENSURE undirected weighted graph $G_u$
\STATE For any given pair of vertices $A,B$ in $G_d$, given the directed signed edges between them, replace them with their corresponding undirected signed weighted version depicted in Figure~\ref{fig:motifs}.  This results in $G_u$, the undirected weighted version of the graph.
$G_u$ may have negative edges. 
\end{algorithmic}
\end{algorithm}

\begin{algorithm}
\caption{{\textbf {\slshape bowler-select}}}
\label{alg:bsel}
\begin{algorithmic}[1]
\REQUIRE undirected weighted graph $G_u$, required bowlers $k$ // ($k$ = 5 or 6.)
\ENSURE $k$ bowlers maximising positive bowlerships 
\STATE	Let C be the set of disconnected components of $G_u$. 
\FOR {each component $c$ in $C$} \label{lst:sfor1}
	\FOR {$i \in \{2,\ldots,k\}$}
		\STATE	Find all subgraphs $S_i = \{S_{i1},\ldots, S_{ip}\}$ of size $i$.
		\STATE	For each subgraph, calculate $W(S_{ij}) = {{\sum w(S_{ij})}\over{|S_{ij}|}}$, \\
		                     where $w(S)$ is the sum of the edge weights induced by $S$. 
	\ENDFOR \label{lst:efor1}
\ENDFOR
\STATE	Output subgraph $S_{max}$ with maximum ${W(S_{ij})} $. \label{lst:op1}
\FOR		{$|S_{ij}|\in \{1,\ldots,(k - |S_{max}|)\}$} \label{lst:sfor2}
	\STATE	Let $X_{ij}$ be the set of cross edges connecting $S_{ij}$ and $S_{max}$, and $w(X_{ij})$ be the sum of the weights of such edges.
	\STATE	$WT(S_{ij}) \leftarrow (W(S_{ij}) + w(X_{ij}))$
\ENDFOR \label{lst:efor2}
\STATE	remain $\leftarrow (k - |S_{max}|).$  \label{lst:srest}
\STATE	size $\leftarrow$ remain. 
\WHILE	{remain $\not=$ 0}
	\IF	{size $>$ remain}
		\STATE size $\leftarrow$ remain
	\ENDIF
	\STATE	Select $S_{ij}$ with $|S_{ij}| = size$ with maximum $WT(S_{ij})$.
	\IF		{no such $S_{ij}$ exists,}
		\STATE	size $\leftarrow$ size - 1.
		\STATE	{\bf{continue}}
	\ENDIF
	\STATE	Output $S_{ij}$. 
	\STATE	remain $\leftarrow remain - |S_{ij}|$.
\ENDWHILE \label{lst:erest}
\end{algorithmic}
\end{algorithm}

The higher the average weighted degree of a subgraph, the more the bowlership synergy among the corresponding bowlers.
Maximising the average weighted degree of the subgraph of the selected bowlers during team (bowler) selection before the match increases the flexibility of bowling changes during the match for the captain. 

Algorithm {\textbf {\slshape bowler-select}} takes as input $G_u$ and a number $k$ of bowlers to be selected and returns a set of bowlers such that the average weighted degree of the selected bowlers is greedily maximised. Steps~\ref{lst:sfor1}--~\ref{lst:efor1} computes the average weighted degree for each {\em connected} subgraph of size $\leq k$, the required number of bowlers. Step~\ref{lst:op1} outputs a subgraph $S_{max}$ with the maximum average weighted degree. If the size of this subgraph equals $k$, then we are done. Otherwise, we need to find another subgraph to fulfill the $k$ bowler requirement. For this, we need to take into account not just the weight of a candidate subgraph in consideration, but also the weight of its connectivity with $S_{max}$. This total weight is calculated in Steps~\ref{lst:sfor2}--~\ref{lst:efor2}. The actual selection of any remaining bowlers is done in Steps~\ref{lst:srest}--~\ref{lst:erest}, until the required number of bowlers $k$ is reached. While there are other algorithms in literature~\cite{charikar2000greedy,tsourakakis2019novel,safonova2015}, some of them supporting negative edge weights, none of them fit our needs exactly. Further, since the number $k$ of required bowlers in our setting is small, the exhaustive computation in steps~~\ref{lst:sfor1}--~\ref{lst:efor1} is practical. 


%
%

\section{Results}

\subsection{Mann-Whitney Analysis}


\begin{figure}
     \centering
          \includegraphics[width=\textwidth]{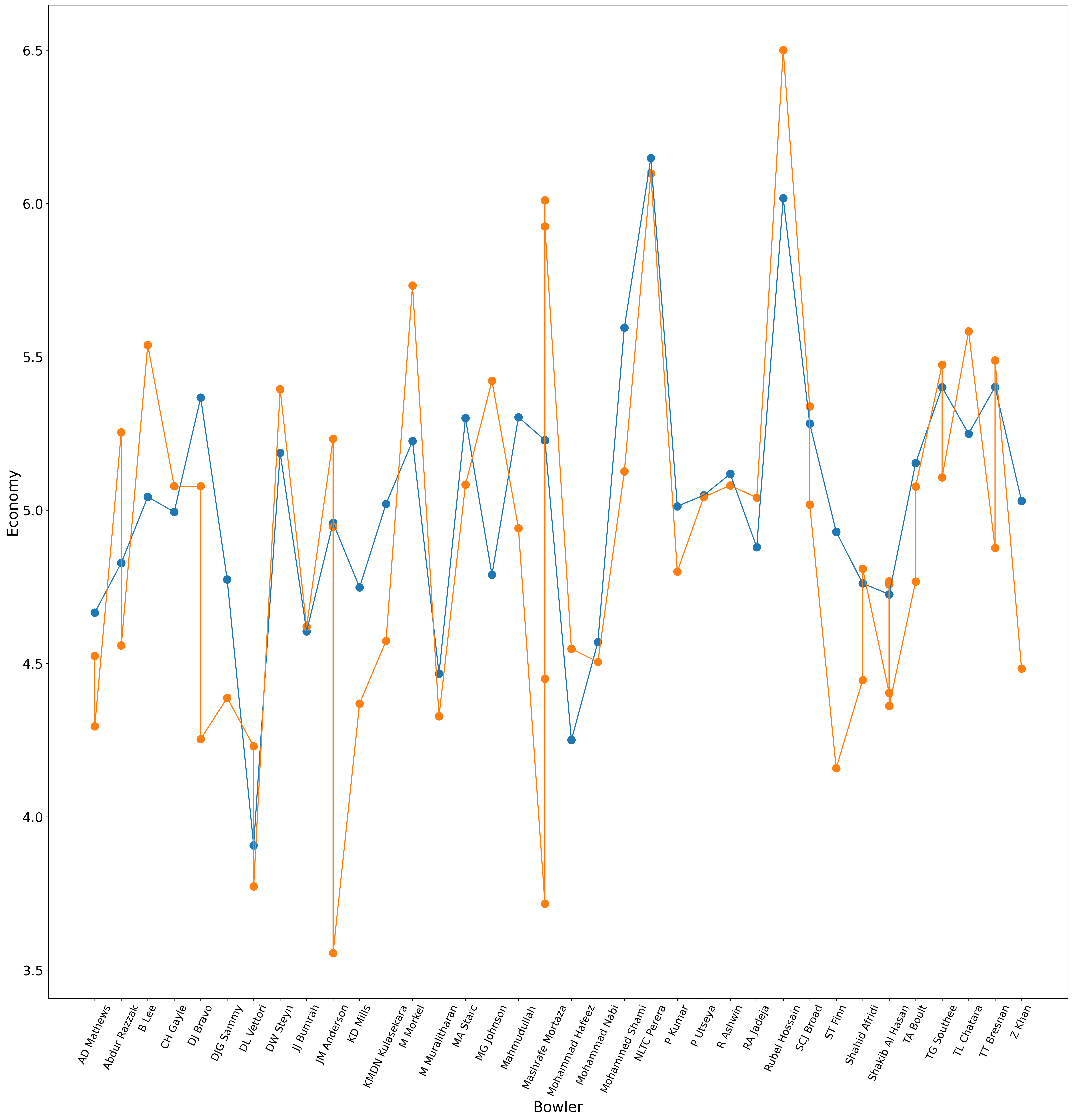}
          \caption{ODI individual and bowlership economy: The X-axis contains the name of the bowler and the Y-axis is the economy. The plot points in orange show the bowlership economy whereas the plot points in blue depict the bowler’s economy with all other bowler partners combined. Note that, the bowlership economy need not be better than the overall economy for a bowlership to be positive.}
      \label{fig:odibshipeco}
\end{figure}
 
\begin{figure}
       \centering
        \includegraphics[width=\textwidth]{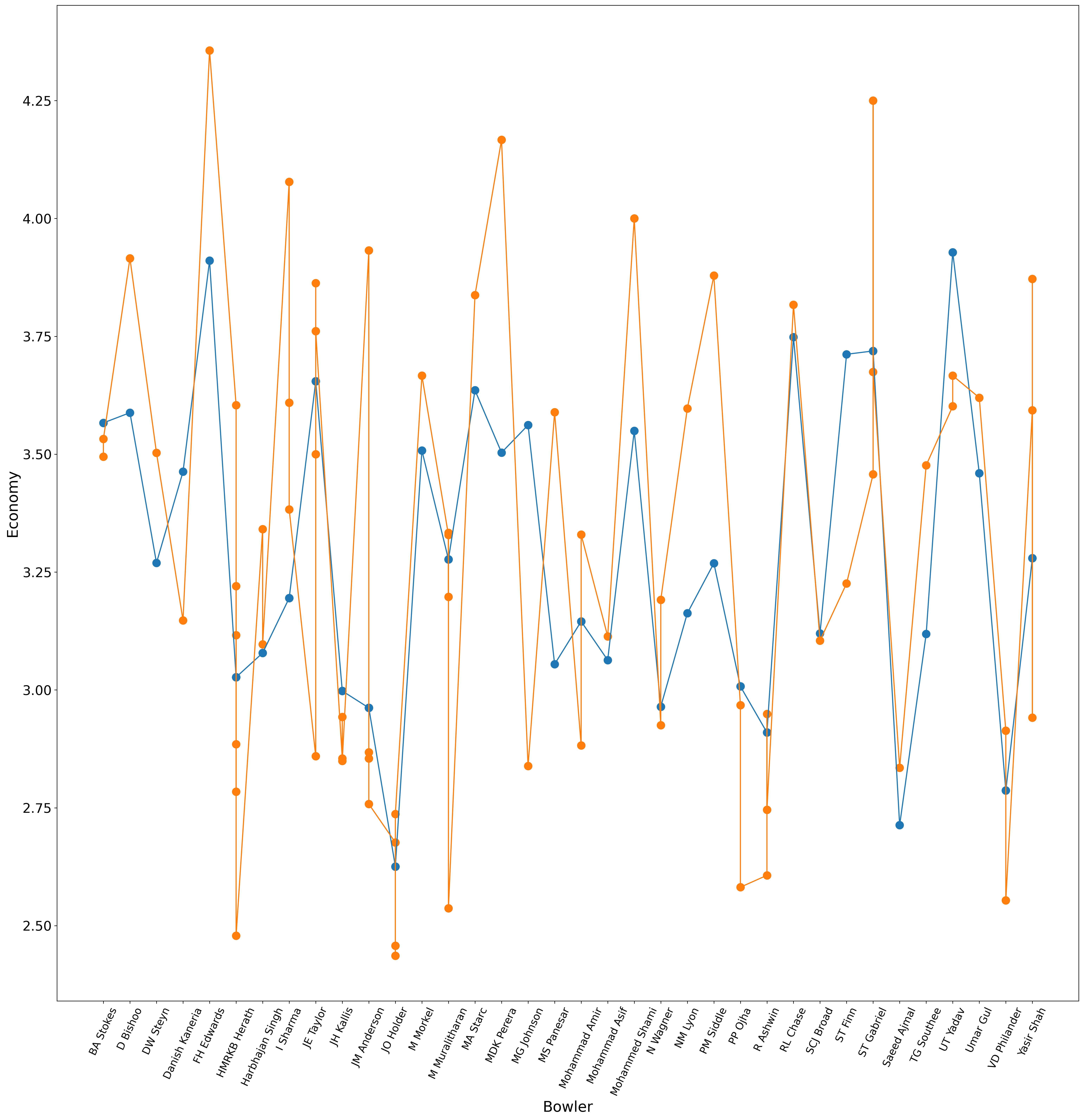}
        \caption{Test individual and bowlership economy: The X-axis contains the name of the bowler and the Y-axis is the economy. The plot points in orange show the bowlership economy whereas the plot points in blue depict the bowler’s economy with all other bowler partners combined. Note that, the bowlership economy need not be better than the overall economy for a bowlership to be positive.}
          \label{fig:testbshipeco}
\end{figure}

\begin{figure}
      \centering
    \includegraphics[width=\textwidth]{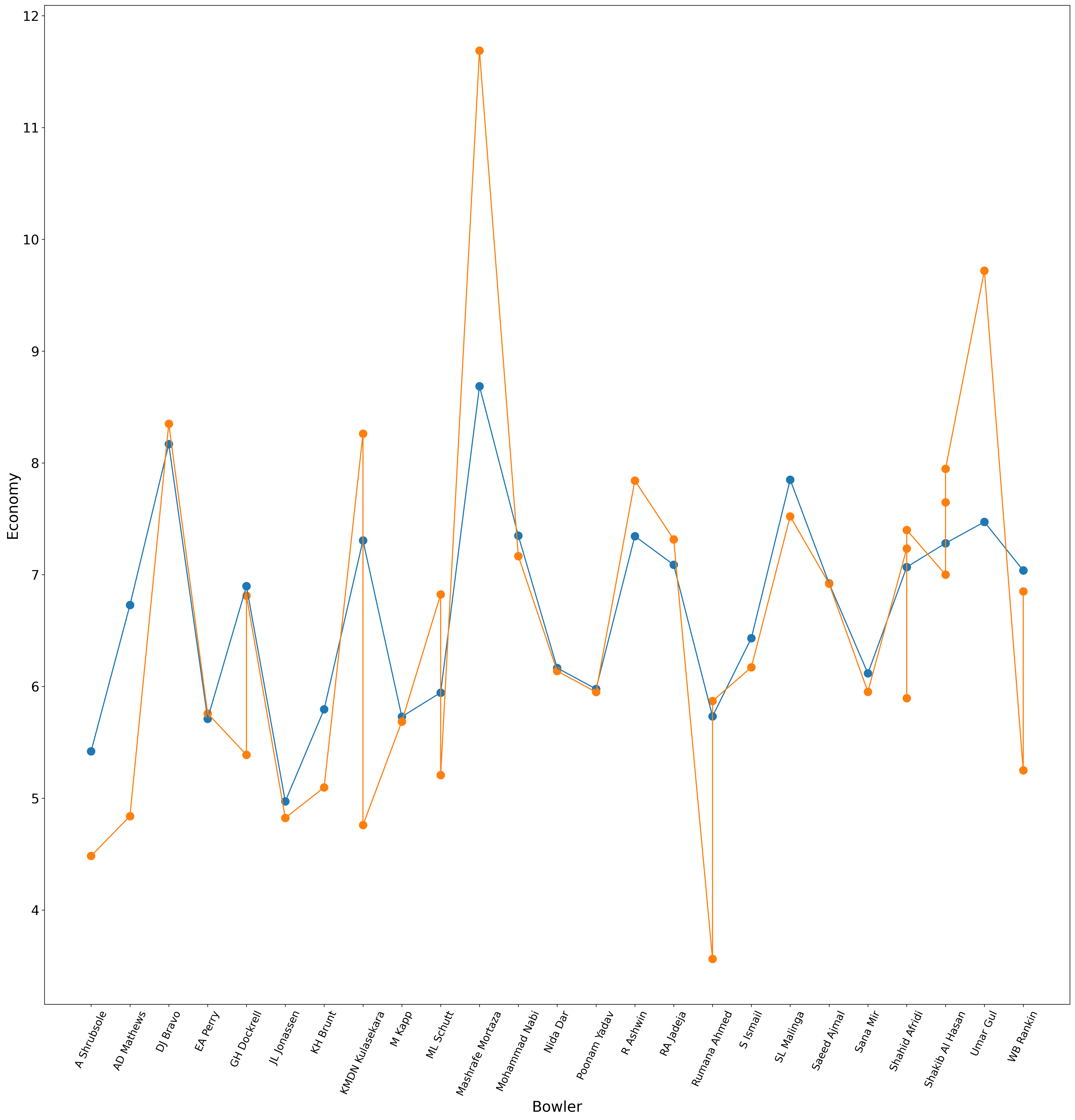}
   \caption{T20I individual and bowlership economy: The X-axis contains the name of the bowler and the Y-axis is the economy. The plot points in orange show the bowlership economy whereas the plot points in blue depict the bowler’s economy with all other bowler partners combined. Note that, the bowlership economy need not be better than the overall economy for a bowlership to be positive.}
    \label{fig:t20bshipeco}
\end{figure}

\begin{figure}
     \centering
           \includegraphics[width=\textwidth]{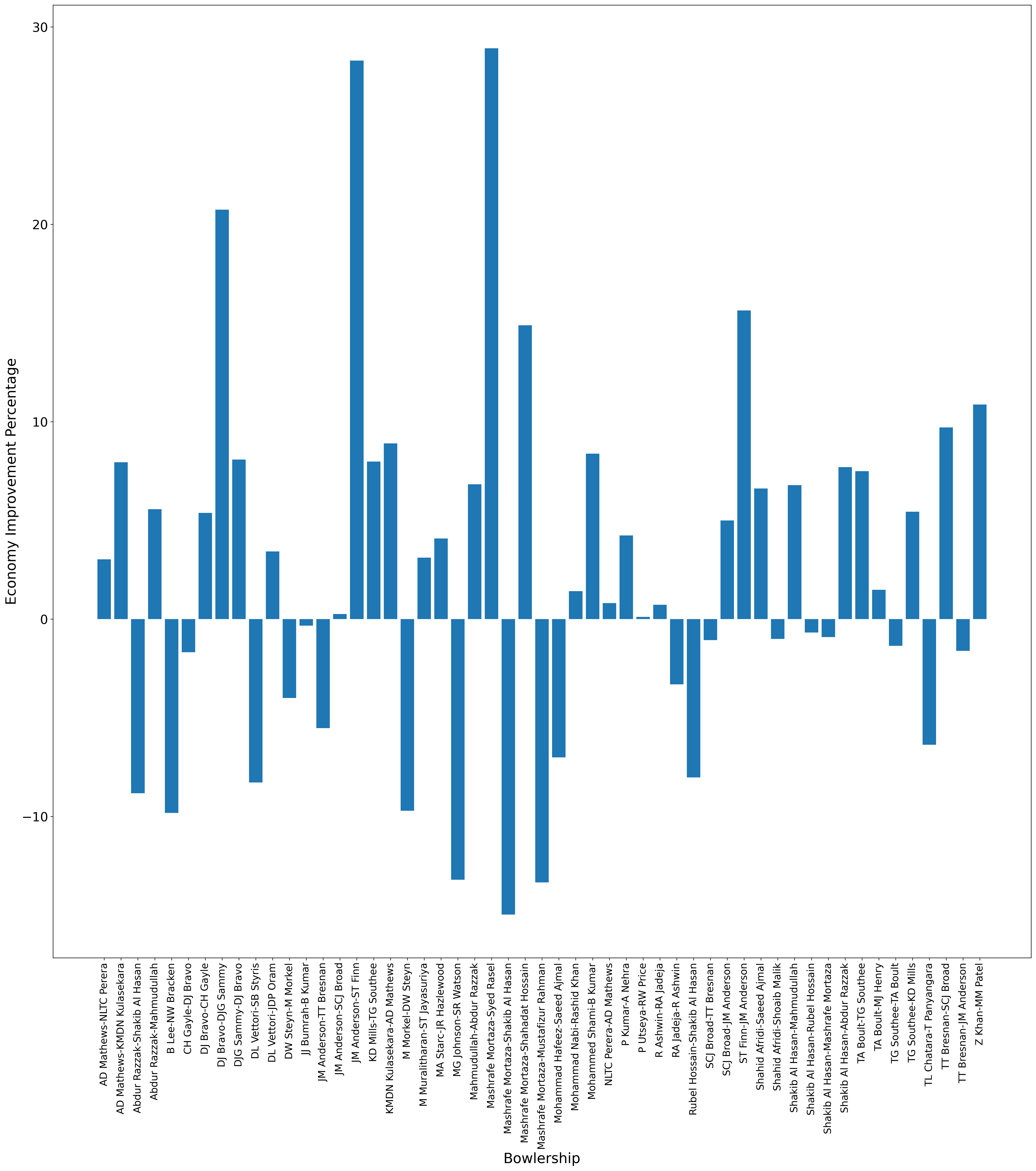}
          \caption{ODI economy improvement in bowlership: This plot highlights the differences in economies of positive bowlerships and individual bowlers.}
          \label{fig:odibshimp}
\end{figure}

\begin{figure}
          \centering
          \includegraphics[width=\textwidth]{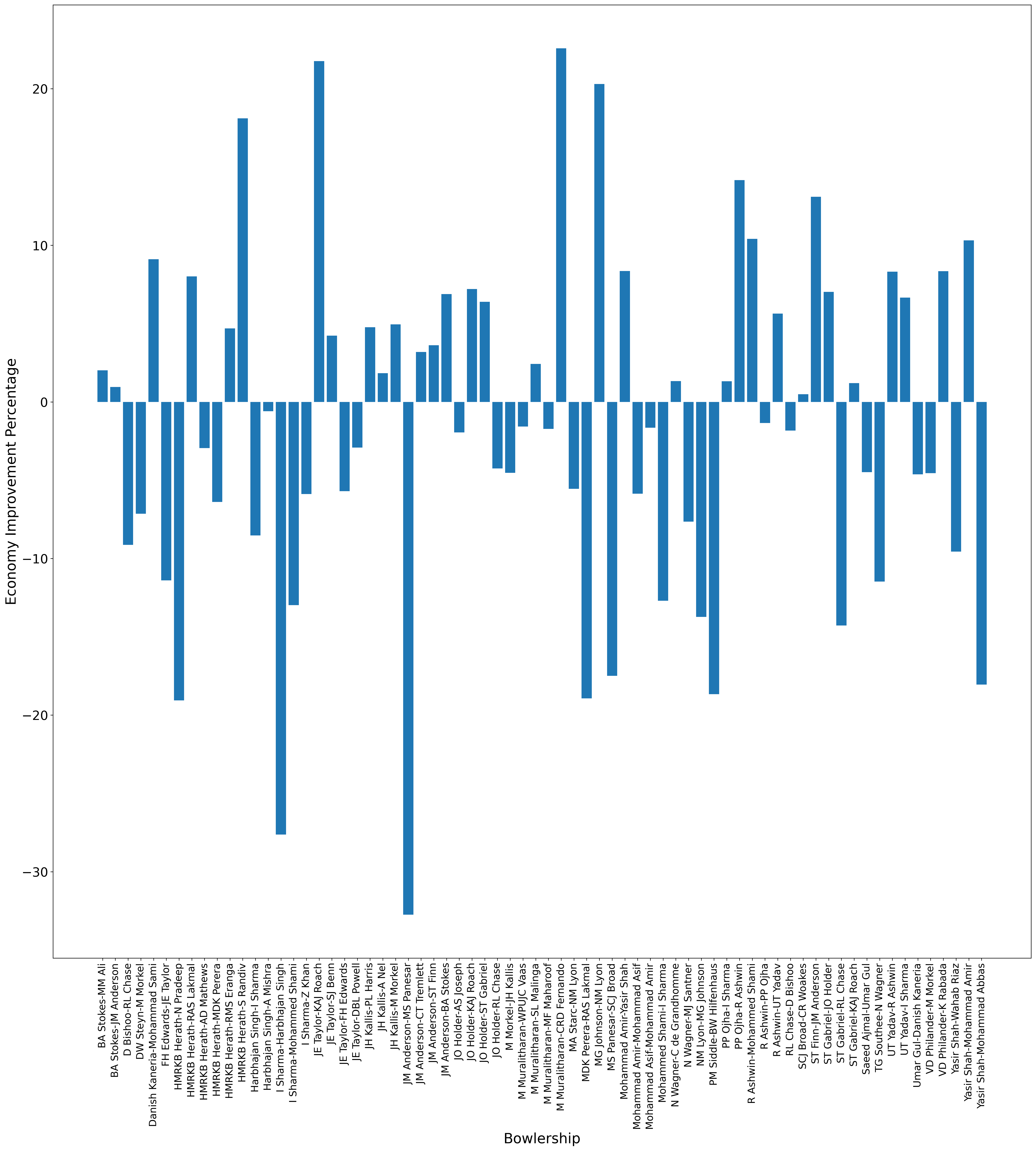}
          \caption{Test economy improvement in bowlership: This figure highlights the differences in economies of positive bowlerships and individual bowlers.}
          \label{fig:testbshimp}
\end{figure}

\begin{figure}
          \centering
          \includegraphics[width=\textwidth]{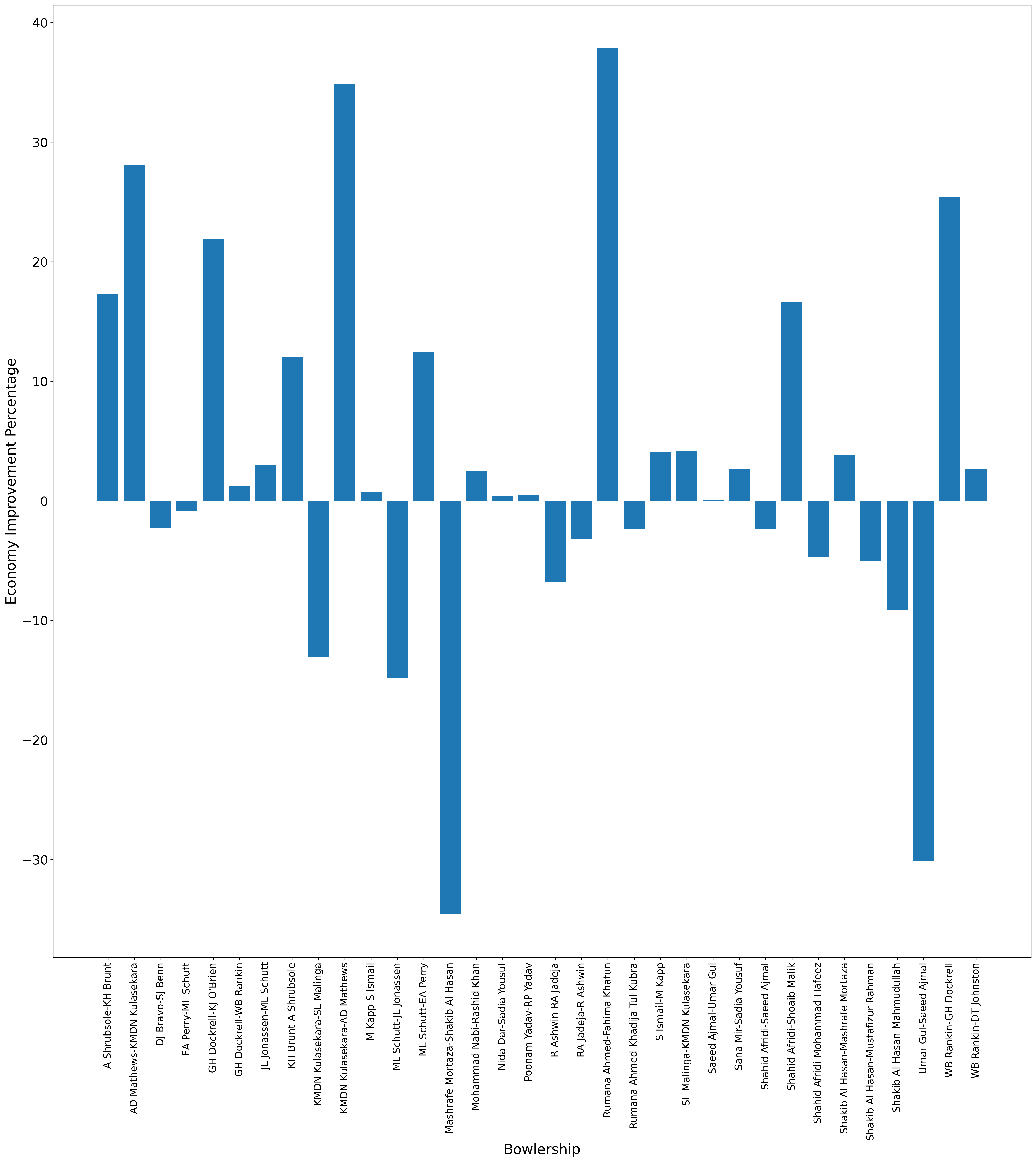}
          \caption{T20I economy improvement in bowlership: This figure highlights the differences in economies of positive bowlerships and individual bowlers.}
          \label{fig:t20bshimp}
\end{figure}

\begin{figure}
     \centering
      \includegraphics[width=\textwidth]{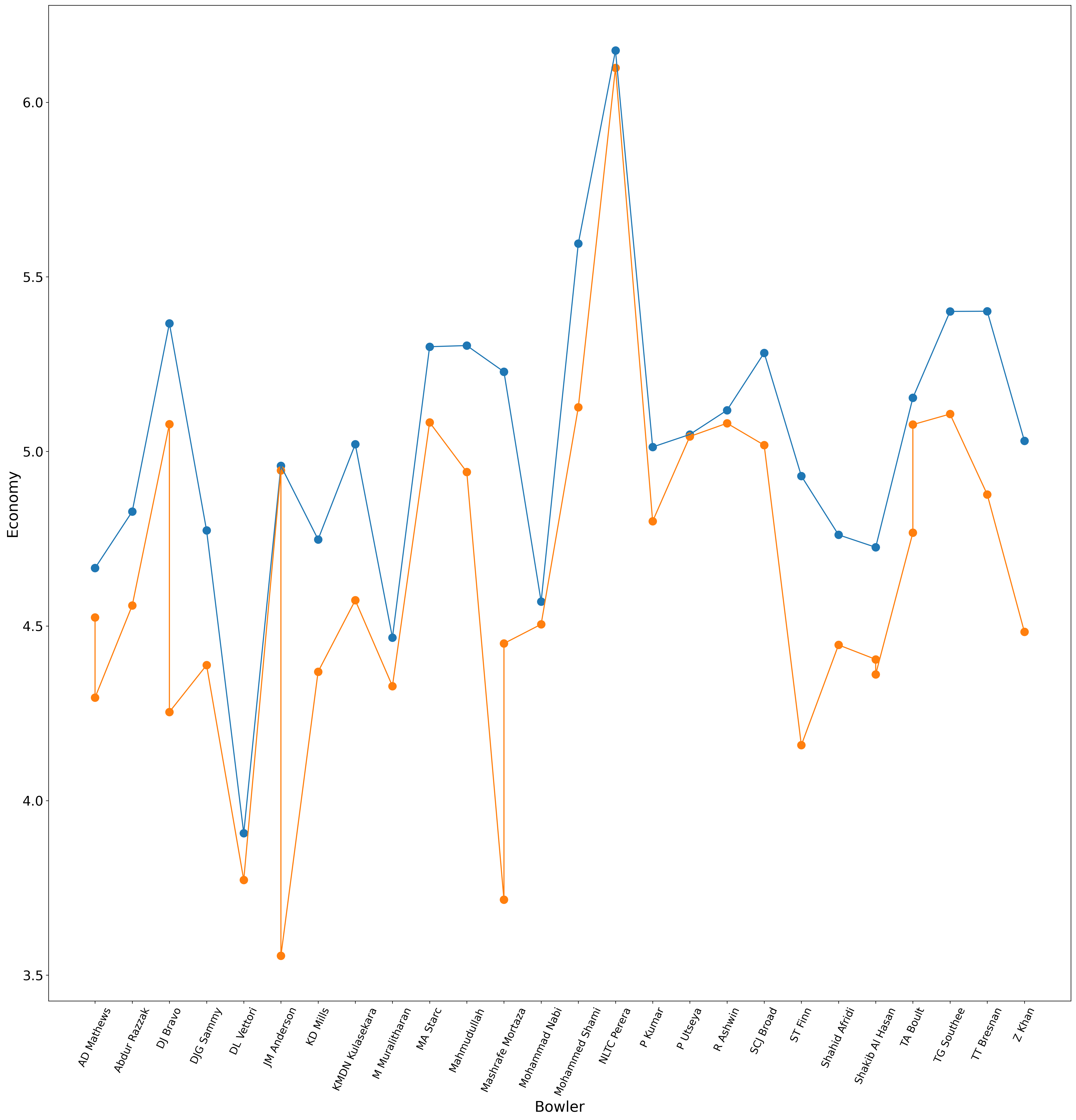}
     \caption{ODI positive economy change: This plot shows only those points where the positive bowlership has a better economy than that of an individual bowler. }
      \label{fig:odibshimpos}
\end{figure}
 
\begin{figure}
      \centering
      \includegraphics[width=\textwidth]{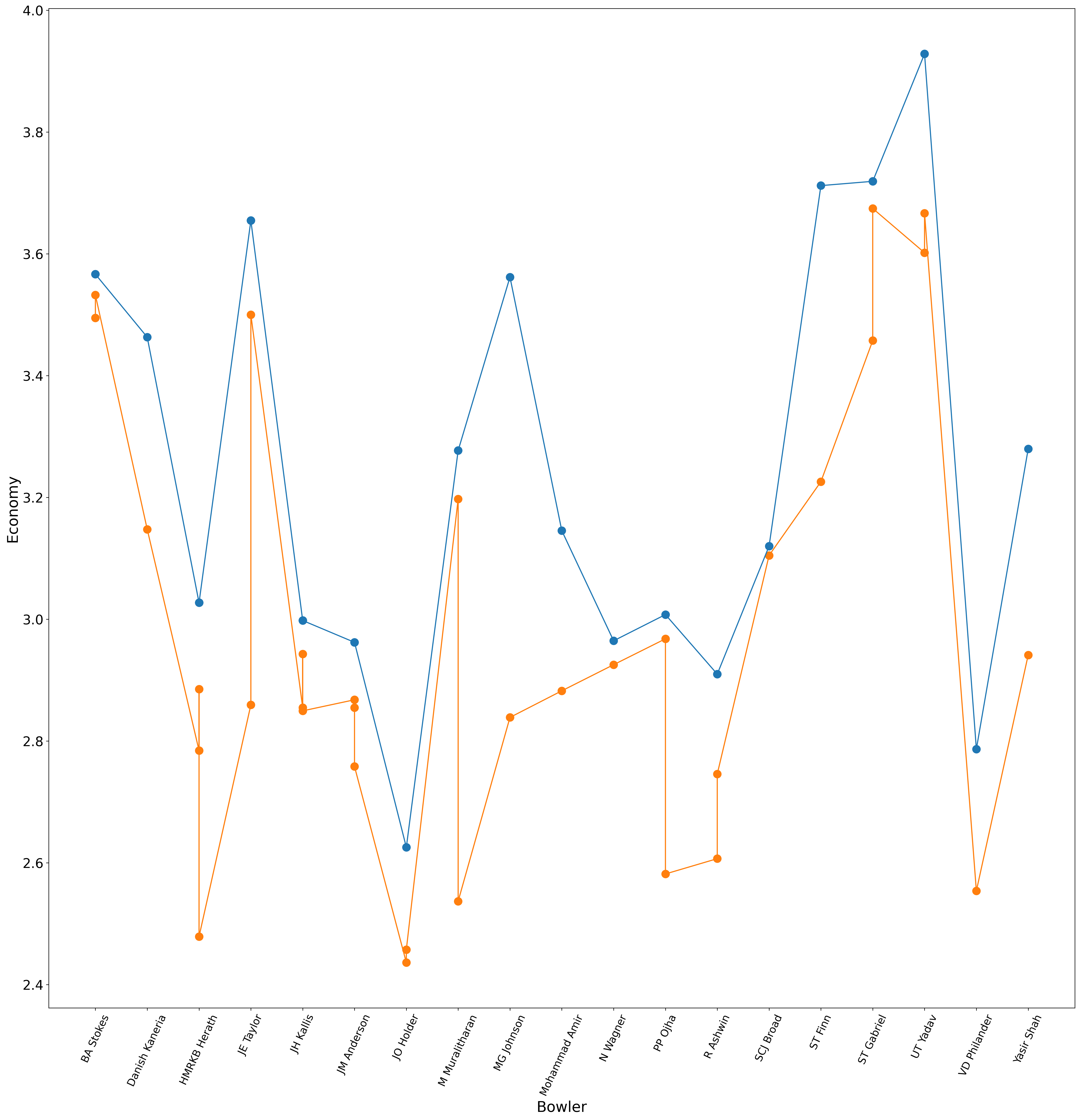}
      \caption{Test positive economy change: This plot shows only those points where the positive bowlership has a better economy than that of an individual bowler.}
      \label{fig:testbshimpos}
\end{figure}

\begin{figure}
     \centering
     \includegraphics[width=\textwidth]{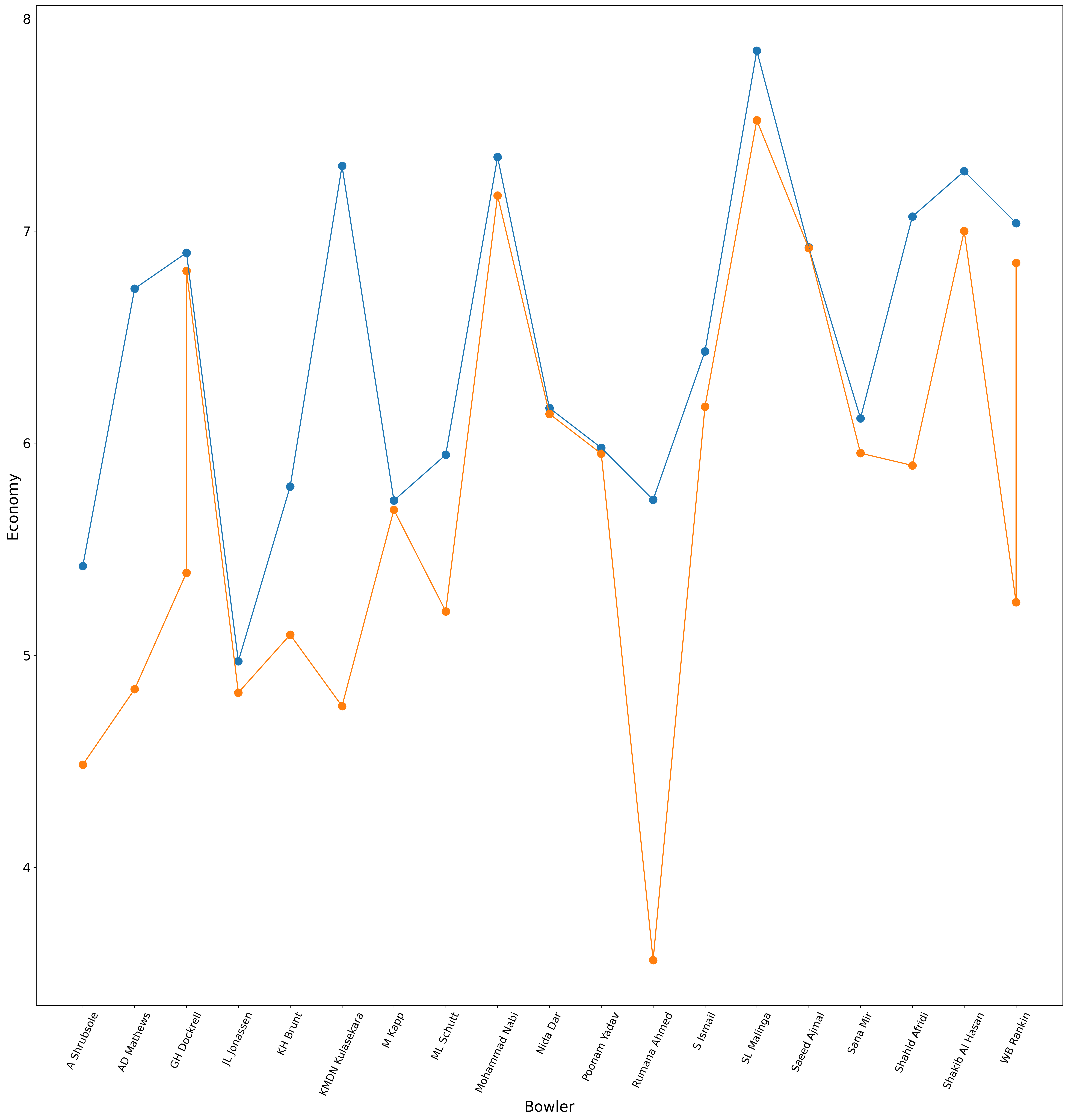}
     \caption{T20I positive economy change: This plot shows only those points where the positive bowlership has a better economy than that of an individual bowler.}
     \label{fig:t20bshimpos}
\end{figure}

\begin{figure}
     \centering
          \includegraphics[width=\textwidth]{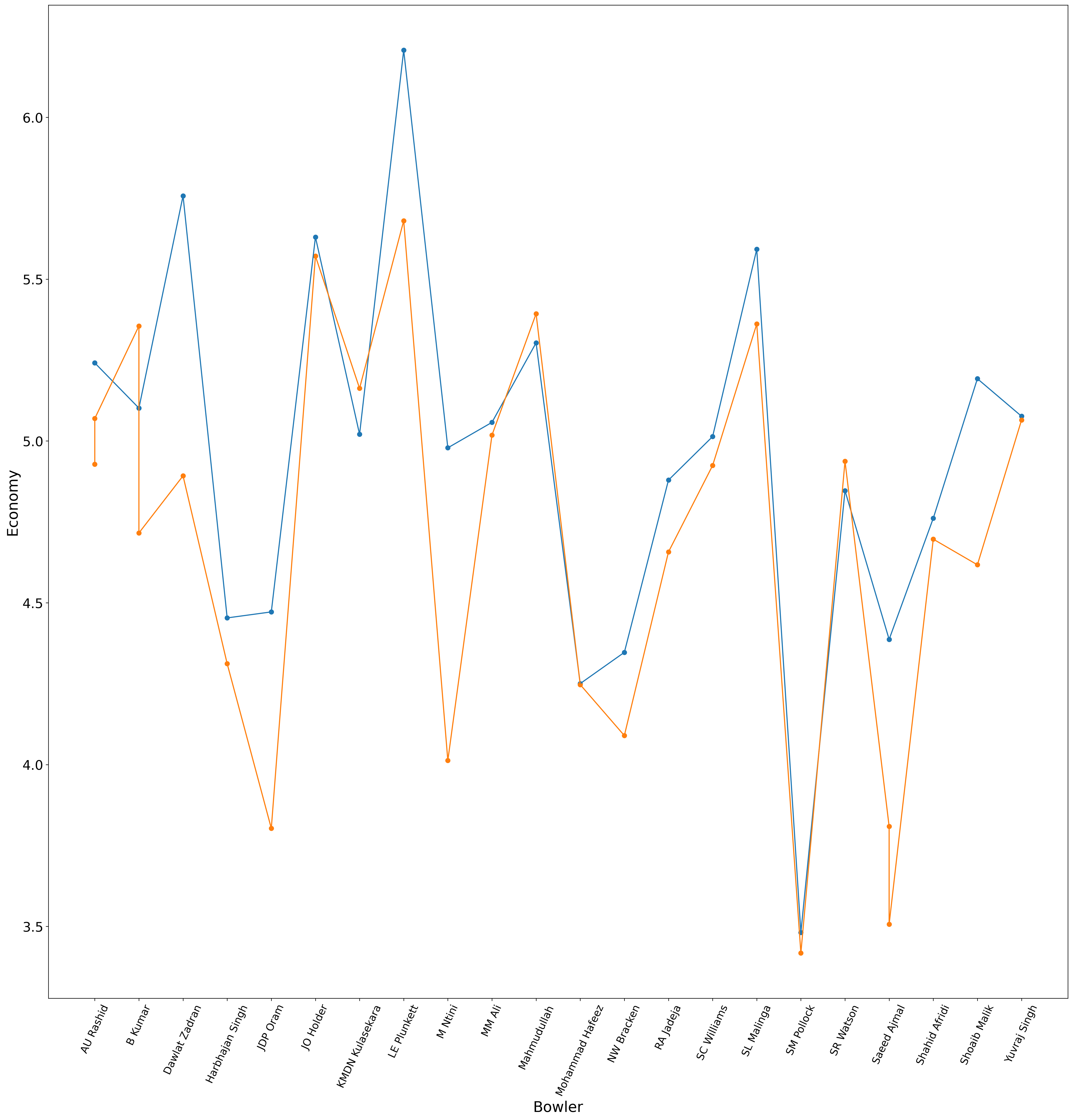}
          \caption{ODI nonpositive bowlerships: This figure shows non-positive bowlership economies in comparison with their individual economies.}
          \label{fig:odinonpos}
\end{figure}

\begin{figure}
          \centering
          \includegraphics[width=\textwidth]{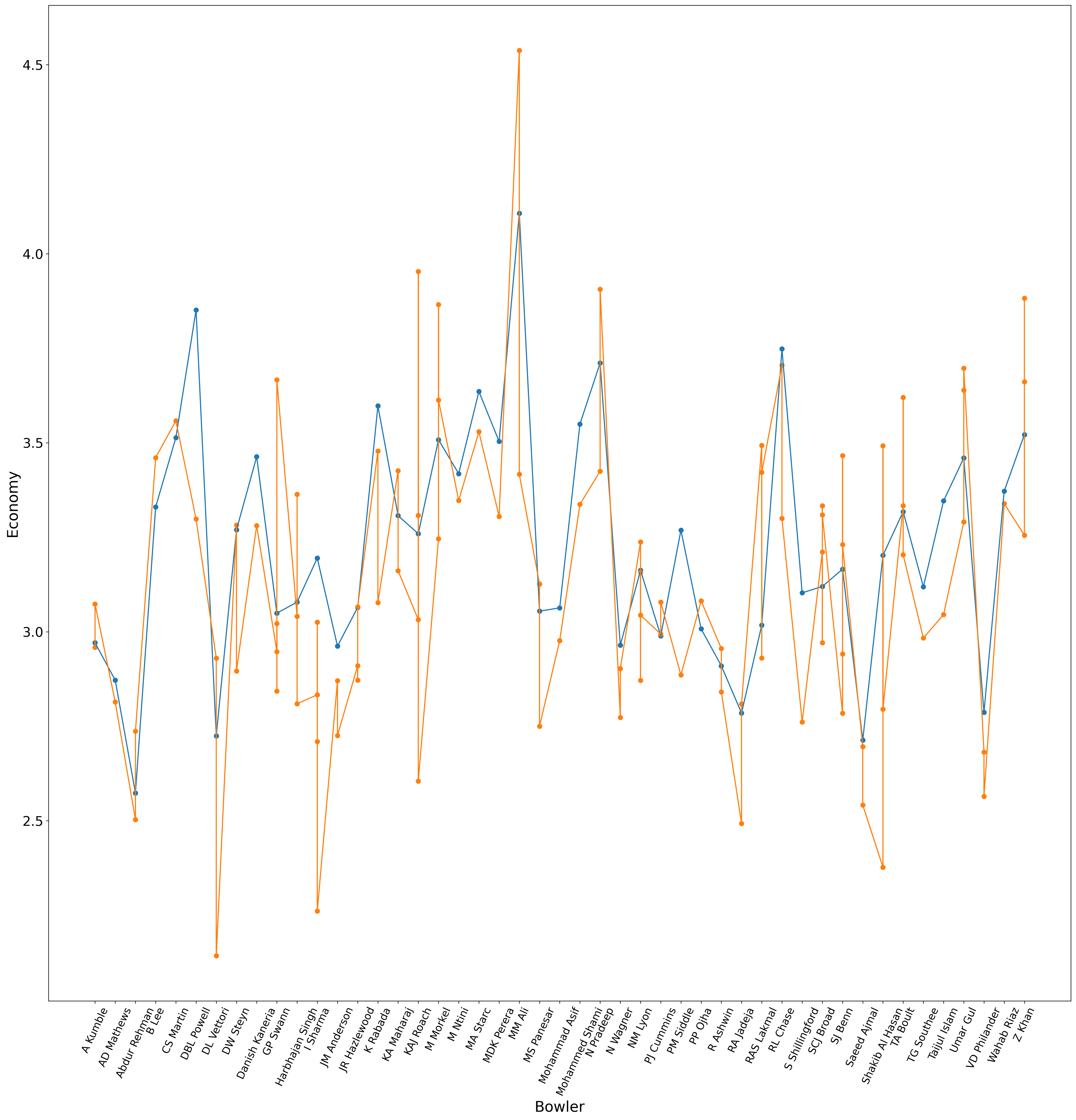}
          \caption{Test nonpositive bowlerships: This figure shows non-positive bowlership economies in comparison with their individual economies.}
          \label{fig:testnonpos}
\end{figure}

\begin{figure}
          \centering
          \includegraphics[width=\textwidth]{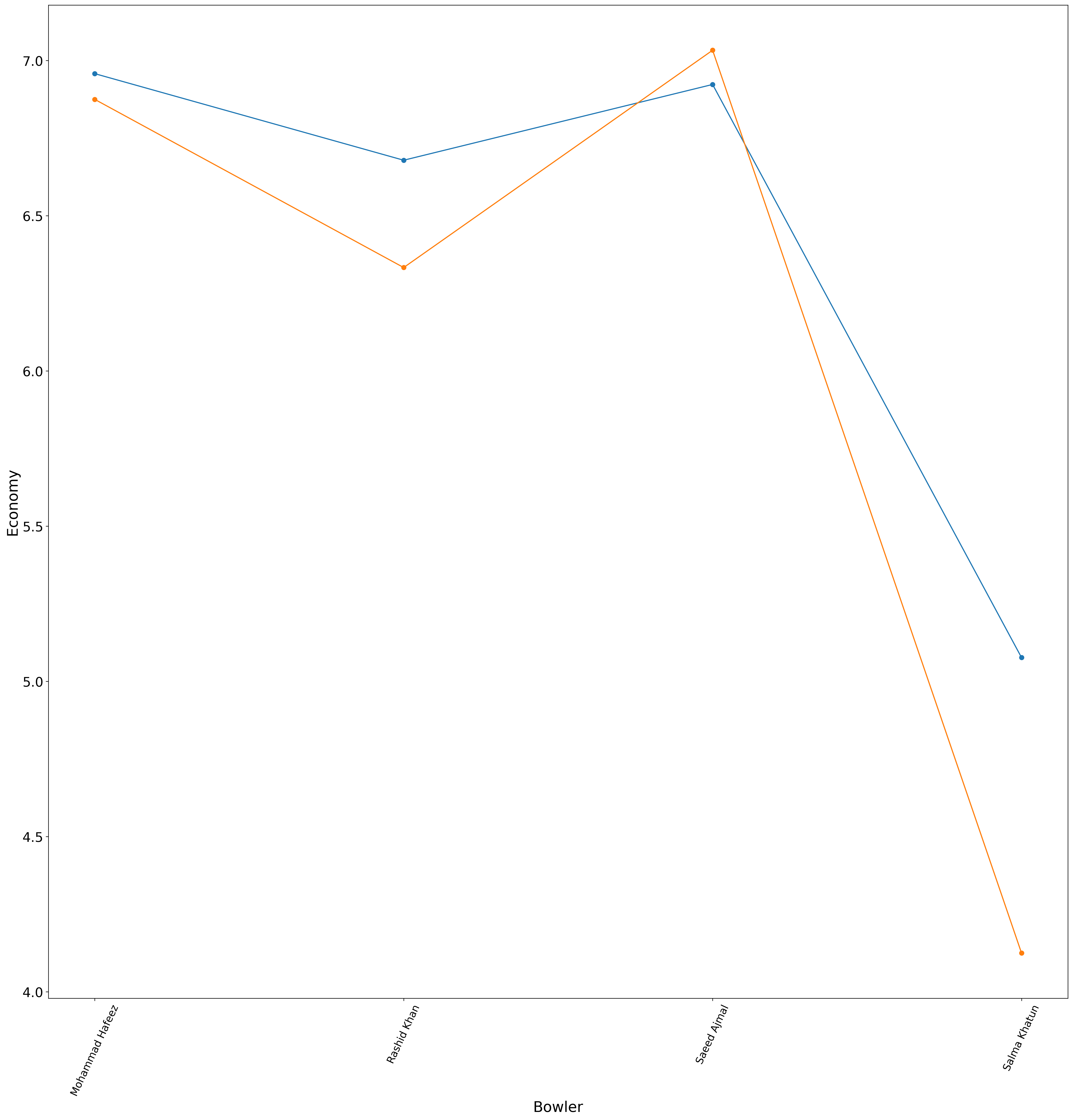}
          \caption{T20I nonpositive bowlerships: This figure shows non-positive bowlership economies in comparison with their individual economies.}
          \label{fig:t20nonpos}
\end{figure}

As we can see in~\Cref{fig:odibshipeco,fig:testbshipeco,fig:t20bshipeco}, the overall economy of a bowler throughout his career in comparison with his overall economy in the bowlership has no role in determining whether the bowlership is `better' than the individual. In these figures, the points in orange depict the bowlership economies, which however, for a majority of bowlers, lie below their individual average economies. This shows that merely taking averages across all overs is not enough.~\Cref{fig:odibshimp,fig:testbshimp,fig:t20bshimp} depict this by showing the difference in economies according to individual bowlers along with their bowlerships and ~\Cref{fig:odibshimpos,fig:testbshimpos,fig:t20bshimpos} shows all positive bowlerships with positive economies.  Another point to note is that economies play a major role mainly in ODIs and T20Is as the runflow is high unlike test matches. Hence, these figures give us more insight about the shorter formats.

Interestingly, we weren't able to reject the null hypotheses of the `two-sided' or the `less' tests for any of the formed bowler pairs for various confidence levels. This means, we did not have any negative bowlerships. Hence, we refer to bowlerships that aren't positive but couldn't be proved negative as non-bowlerships (no edge exists between the pair of bowlers).

As depicted in~\Cref{fig:odinonpos,fig:testnonpos,fig:t20nonpos}, non-bowerships also show similar trends in economies as positive bowlerships.
Similarly, we tried to do the Mann-Whitney experiments with wickets taken per over for bowlers who had bowled at least a certain number of overs. No bowlership turned out to be positive for a range of confidence levels and we were not able to reject any of the null hypotheses. We believe the reason for this is that the taking of a wicket is a rare event occurring only a few times (if at all) per match, in comparison with the number of overs bowled. However, it is possible that bowlerships affect the falling of wickets, but this is statistically insignificant.

\subsection{Bowlership Networks}

\begin{figure}
    \centering
         \begin{subfigure}[b]{0.25\textwidth}
         \centering
         \includegraphics[width=\textwidth]{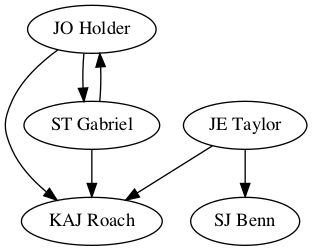}
         \caption{West Indies}
         \label{sfig:test-wi-mw-eco}
         \end{subfigure}
         \hfill
         \begin{subfigure}[b]{0.39\textwidth}
         \centering
         \includegraphics[width=\textwidth]{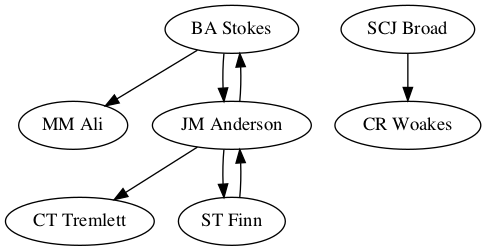}
         \caption{England}
         \label{sfig:test-eng-mw-eco}
         \end{subfigure}
         \hfill 
         \begin{subfigure}[b]{0.33\textwidth}
         \centering
         \includegraphics[width=\textwidth]{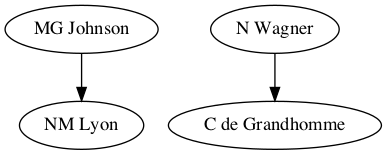}
         \caption{Australia \& New Zealand}
         \label{sfig:test-anz-mw-eco}
         \end{subfigure}
         \hfill \\
        \begin{subfigure}[b]{0.34\textwidth}
         \centering
         \includegraphics[width=\textwidth]{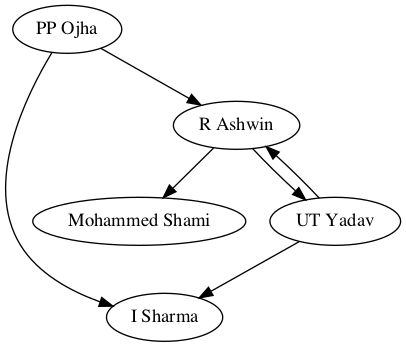}
         \caption{India}
         \label{sfig:test-ind-mw-eco}
         \end{subfigure}
         \hfill
         \begin{subfigure}[b]{0.185\textwidth}
         \centering
         \includegraphics[width=\textwidth]{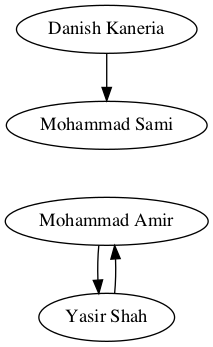}
         \caption{Pakistan}
         \label{sfig:test-pak-mw-eco}
         \end{subfigure}
         \hfill
         \begin{subfigure}[b]{0.34\textwidth}
         \centering
         \includegraphics[width=\textwidth]{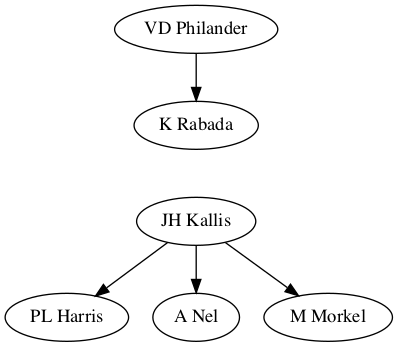}
         \caption{South Africa}
         \label{sfig:test-sa-mw-eco}
         \end{subfigure}
         \hfill \\
         \begin{subfigure}[b]{0.70\textwidth}
         \centering
         \includegraphics[width=\textwidth]{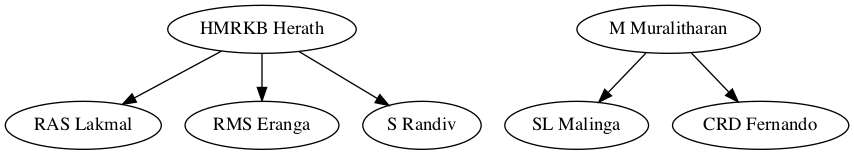}
         \caption{Sri Lanka}
         \label{sfig:test-sl-mw-eco}
         \end{subfigure}
         \hfill \\
    \caption{Test Bowlership Network.}
   \label{fig:test-nw}
\end{figure}

\begin{figure}
    \centering
         \begin{subfigure}[b]{0.23\textwidth}
         \centering
         \includegraphics[width=\textwidth]{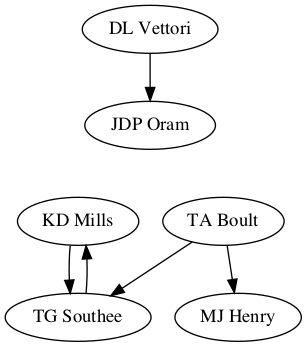}
         \caption{New Zealand}
         \label{sfig:odi-nz-mw-eco}
         \end{subfigure}
         \hfill
         \begin{subfigure}[b]{0.26\textwidth}
         \centering
         \includegraphics[width=\textwidth]{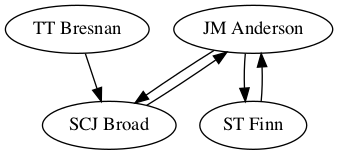}
         \caption{England}
         \label{sfig:odi-eng-mw-eco}
         \end{subfigure}
         \hfill 
         \begin{subfigure}[b]{0.48\textwidth}
         \centering
         \includegraphics[width=\textwidth]{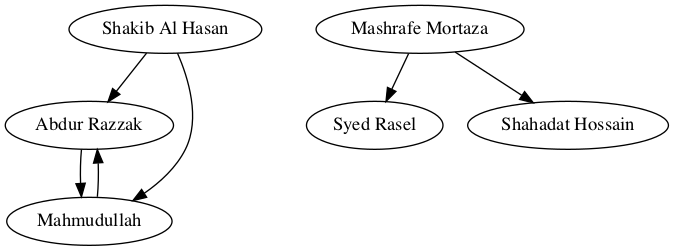}
         \caption{Bangladesh}
         \label{sfig:odi-ban-mw-eco}
         \end{subfigure}
         \hfill \\
        \begin{subfigure}[b]{0.23\textwidth}
         \centering
         \includegraphics[width=\textwidth]{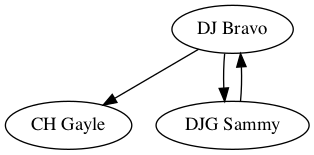}
         \caption{West Indies}
         \label{sfig:odi-wi-mw-eco}
         \end{subfigure}
         \hfill
         \begin{subfigure}[b]{0.28\textwidth}
         \centering
         \includegraphics[width=\textwidth]{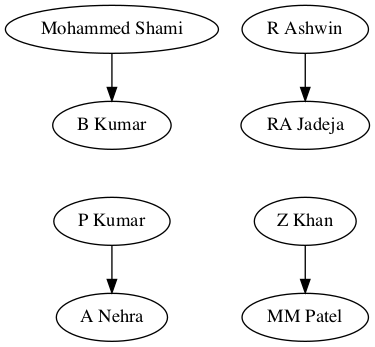}
         \caption{India}
         \label{sfig:odi-ind-mw-eco}
         \end{subfigure}
         \hfill
         \begin{subfigure}[b]{0.44\textwidth}
         \centering
         \includegraphics[width=\textwidth]{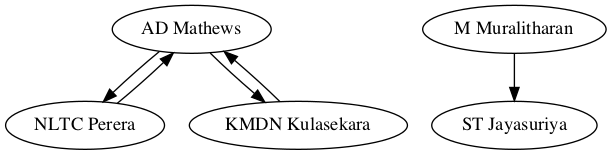}
         \caption{Sri Lanka}
         \label{sfig:odi-sl-mw-eco}
         \end{subfigure}
         \hfill \\
         \begin{subfigure}[b]{0.58\textwidth}
         \centering
         \includegraphics[width=\textwidth]{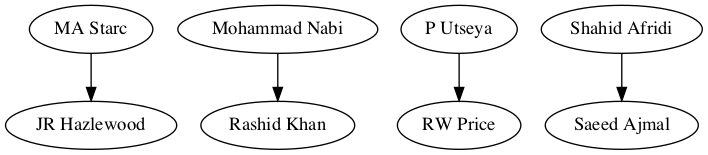}
         \caption{Australia, Afghanistan, South Africa \& Pakistan}
         \label{sfig:odi-sl-mw-eco}
         \end{subfigure}
         \hfill \\
    \caption{ODI Bowlership Network}
   \label{fig:odi-nw}
\end{figure}

\begin{figure}
    \centering
         \begin{subfigure}[b]{0.30\textwidth}
         \centering
         \includegraphics[width=\textwidth]{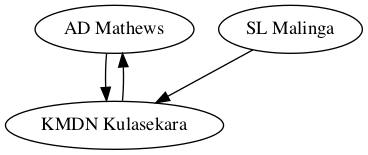}
         \caption{Sri Lanka}
         \label{sfig:t20-sl-mw-eco}
         \end{subfigure}
         \hfill
         \begin{subfigure}[b]{0.25\textwidth}
         \centering
         \includegraphics[width=\textwidth]{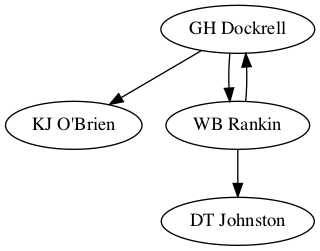}
         \caption{Ireland}
         \label{sfig:odi-ire-mw-eco}
         \end{subfigure}
         \hfill 
         \begin{subfigure}[b]{0.30\textwidth}
         \centering
         \includegraphics[width=\textwidth]{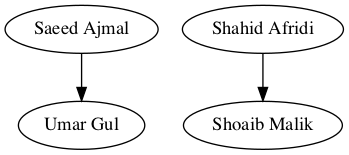}
         \caption{Pakistan}
         \label{sfig:t20-pak-mw-eco}
         \end{subfigure}
         \hfill \\
        \begin{subfigure}[b]{0.37\textwidth}
         \centering
         \includegraphics[width=\textwidth]{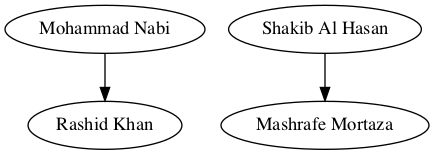}
         \caption{Afghanistan \& Bangladesh}
         \label{sfig:odi-wi-mw-eco}
         \end{subfigure}
         \hfill \\
    \caption{T20 Bowlership Network}
   \label{fig:t20-nw}
\end{figure}

Figures~\ref{fig:test-nw}, ~\ref{fig:odi-nw} and ~\ref{fig:t20-nw} depict the bowlerships networks for Tests, ODIs and T20Is respectively. Some interesting points to be noted are:

\begin{itemize}
\item The test network is the densest and the T20 Network is
sparsest. This could be due to frequent bowling changes in the shorter
formats as opposed to tests.

\item While it is an artefact of the chosen parameters (T1 and T2), two way positive bowlerships are quite uncommon across all formats.  We also expected to see fewer disconnected components.

\item For the Test Bowlership Network~\ref{fig:test-nw}, running
Algorithm {\textbf {\slshape bowler-select}} for West Indies and $k=3$, results in
\{J.O. Holder, S.T. Gabriel, K.A.J. Roach\} getting selected. For England and $k=4$,
the algorithm selects \{B.A. Stokes, J.M. Anderson, S.T. Finn\} and either one of 
M.M. Ali and C.T.Tremlett. These results look quite intuitive. 
In practice, Algorithm {\textbf {\slshape bowler-select}} may run efficiently since the sizes of these graphs is small. 

\item We expected to see a few cycles of size 3 or more, but did not. This 
could again be an artefact of the chosen parameters (T1 and T2).

\end{itemize}

\section{Discussion}
\label{sec:discuss}

A positive bowlership does not necessarily imply better economy. In practice, we could select positive bowlerships that also have better economies. The analysis with wickets did not yield any conclusive results, as the falling of a wicket is a rare event.

For the threshold values chosen in the experiments, the bowlership networks have disconnected components, with no components having more than 5 vertices. Such networks can therefore be visually analysed. We also saw that Algorithm {\textbf{\slshape bowler-select}} does give intuitive results.

The more the outgoing edges from a bowler, the more the number of bowlers he bowls better with. 
The more the more incoming edges to a bowler,the more the number of bowlers who bowl better with him.
Therefore, the number of incident edges (incoming or outgoing) is an important parameter for picking a bowler.

For example, in Sri Lankan ODIs, selecting A.D. Mathews, N.L.T.C. Perera and K.M.D.N. Kulasekara
together and planning their bowling attacks carefully can prove to be helpful
as Perera and Kulasekara both perform better with Mathews and
vice versa. Similarly, S.T. Finn, J.M. Anderson and S.C.J. Broad could be picked for
England ODIs.

For tests, South Africans must pick Jacques Kallis and Sri Lankans H.M.R.K.B. Herath as
they are very compatible with three other bowlers and seem to perform well with
them. Similarly, West Indies should try and include K.A.J. Roach in their eleven while
playing tests as J.O. Holder, J.E. Taylor and S.T. Gabriel give away fewer runs when
bowling with him.

\section{Conclusion}
\label{sec:concl}

Player synergies are an integral and crucial part of team sports.  Unlike
games such as football and tennis, where all the players are participating in
the same activity at the same time, cricket and baseball have one team fielding
while the other is batting. In cricket, batting partnerships have been analysed
extensively, bowling partnerships have not. We first define what constitutes
a bowling partnership, and then analyse all formats of the game in search of
effective bowler partnerships. i.e., ``bowlerships".

We analysed ball-by-ball data for 2,034 ODIs, 634 Test matches and 1,432 T20Is.
There were a total of 1148 ODI bowlers, 495 test bowlers and 1518 T20I bowlers.
For the analysis, we chose the following thresholds: (i) Each individual bowler
in a bowling pair should have bowled at least 300 overs (in Tests), 300 overs
(in ODIs) and 80 overs (in T20Is) throughout the span of their careers. (ii)
In order for a pair of bowlers to be considered a bowling pair, we set the
pairing-threshold -- the number of consecutive overs that they should have
bowled alternately -- to 60 (in Tests), 60 (in ODIs) and 16 (in T20Is).

Our analyses showed that bowlerships exist. These bowlerships can be 
leverages both strategically (for team formation) and tactically (for bowling
changes while the match is in progress). We presented Algorithm 
{\textbf{\slshape bowler-select}} to select bowlers which account for 
bowler synergies during team selection.

In future work, it would be very interesting to investigate the various 
bowlership patterns that emerge based on varying the thresholds. Is there a
systematic way to determine the thresholds? Also, we need to look deeper into
the reason why negative bowlerships were not found. 

How do we compare a pair of positive bowlerships? Can we add weights to the directed 
signed graph $G_d$? This would help us differentiate between stronger and weaker bowlership
pairs. 

\bibliography{bowlership-base}

\end{document}